\ifdefined\arXiv
\documentclass{article}
\pdfoutput=1
\usepackage[margin=.8in]{geometry}
\usepackage{authblk}
\usepackage[pdftex,letterpaper]{geometry} 
\else
\documentclass[sigconf, nonacm]{acmart}

\fi

\usepackage{pifont}
\newcommand{\cmark}{\ding{51}}%
\newcommand{\xmark}{\ding{55}}%

\newcommand{\titletext}{\SSwRwS: Combining Sketching and Sampling Is Better than Either for Per-item Quantile Estimation}

\usepackage{epsfig}
\usepackage{amsmath}
\usepackage{amsfonts}
\usepackage{graphicx}
\usepackage{balance}  
\usepackage{cleveref}

\usepackage{algpseudocode}
\usepackage{bm}
\usepackage{graphicx}
\usepackage{comment}
\usepackage{array}
\usepackage{float}
\floatstyle{plaintop}
\restylefloat{table}
\ifdefined\VLDB
\newcommand{\subparagraph}{}
\fi
\let\proof\relax
\let\endproof\relax
\usepackage{amsthm}
\usepackage{amsmath}
\usepackage{algorithm}
\usepackage{subfig}
\usepackage{titlesec}
\usepackage{blindtext}
\usepackage{multirow}
\usepackage{yfonts}
\usepackage{xcolor}
\usepackage{url}
\usepackage{balance}
\usepackage{nopageno}
\usepackage[font=small,skip=7pt]{caption}
\usepackage[tableposition=top]{caption}
\usepackage[labelfont=bf]{caption}

\newcommand{\changed}[1]{\textcolor{blue}{#1}} 

\newcommand*{\inlineequation}[2][]{%
	\begingroup
	\refstepcounter{equation}%
	\ifx\\#1\\%
	\else
	\label{#1}%
	\fi
	\relpenalty=10000 %
	\binoppenalty=10000 %
	\ensuremath{%
		#2%
	}%
	~\@eqnnum
	\endgroup
}
\algtext*{EndFunction}
\algtext*{EndIf}
\algtext*{EndFor}
\algtext*{EndWhile}
\algtext*{EndIf}

\newif\ifcomm
\commtrue
\ifcomm
\else
\commfalse
\fi
\ifcomm
	\newcommand{\mycomm}[3]{{\footnotesize{{\color{#2} \textbf{[#1: #3]}}}}}
	\newcommand{\CRdel}[1]{\textcolor{red}{\sout{#1}}}
\else
    \newcommand{\mycomm}[3]{}
    \newcommand{\CRdel}[1]{}
\fi
\newcommand{\roy}[1]{\mycomm{roy}{purple}{#1}} 
\newcommand{\ran}[1]{\mycomm{Ran}{blue}{#1}} 
\newcommand{\rana}[1]{\mycomm{Rana}{red}{#1}}

\newtheorem{theorem}{Theorem}

\newcommand{\HHLproblem}{$(\theta,\epsilon,\delta)$-HH-latencies problem}
\newcommand{\HHLproblemParams}[3]{$(#1,#2,#3)$-HH-latencies problem}

\newcommand{\RwS}{SQUARE\xspace}
\newcommand{\RwSFULL}{Sampled QUAntile REconstruction}
\newcommand{\SSwGK}{QUASI\xspace}
\newcommand{\SSwGKFULL}{QUAntile Sketches for heavy Items}
\newcommand{\SSwRwS}{SQUAD\xspace}
\newcommand{\SSwRwSFULL}{Sketching/sampling QUAntiles Duo}

\ifdefined\EXTENDED
\newcommand{\matrixCellWidth}{5.8cm}
\else
\newcommand{\matrixCellWidth}{5.5cm}
\fi

\renewcommand{\mid}{\ensuremath{:}}

\newcommand{\eps}{\epsilon}
\newcommand{\brackets}[1]{\ensuremath{\left[#1\right]}}

\newcommand{\set}[1]{\left\{#1\right\}}
\newcommand{\ceil}[1]{ \left\lceil{#1}\right\rceil}

\newcommand{\angles}[1]{ \left\langle{#1}\right\rangle}

\newcommand{\parentheses}[1]{ \left({#1}\right)}
\newcommand{\abs}[1]{ \left|{#1}\right|}

\newcommand{\oneOverE}{ \ensuremath{\eps^{-1}} }

\newcommand{\window}{W}

\newcommand{\weps}{\window\epsilon}

\begin{document}



\title{\titletext}

\author{Rana Shahout}
\affiliation{%
Technion
}
\email{ranas@cs.technion.ac.il}

\author{Roy Friedman}
\affiliation{%
Technion
}
\email{roy@cs.technion.ac.il}

\author{Ran Ben Basat}
\affiliation{%
  \institution{University College London}
}
\email{r.benbasat@cs.ucl.ac.uk}

\ifdefined\arXiv
\author{Rana Shahout \qquad Roy Friedman \qquad Ran Ben Basat\\
    Computer Science, Technion\qquad{}\texttt{\{ranas,roy,sran\}@cs.technion.ac.il}}
\author[1]{Rana Shahout \thanks{ranas@cs.technion.ac.il}}
\author[1]{Roy Friedman\thanks{roy@cs.technion.ac.il}}
\author[2]{Ran Ben Basat\thanks{ran@seas.harvard.edu}}

\affil[2]{Computer Science, Harvard University}
\affil[1]{Computer Science, Technion}
\else

%
%
\fi

\setlength{\textfloatsep}{14pt}
\setlength{\intextsep}{16pt}
\setlength{\dbltextfloatsep}{12pt}
\setlength{\abovedisplayskip}{3pt}
\setlength{\belowdisplayskip}{3pt}
\titlespacing*{\section} {0pt}{1.2ex plus .7ex minus .2ex}{1.2ex plus .2ex}
\titlespacing*{\subsection} {0pt}{1.1ex plus .7ex minus .2ex}{1.00ex plus .2ex}
\titlespacing*{\subsubsection}{0pt}{0.9ex plus .7ex minus .2ex}{0.90ex plus .2ex}

\begin{abstract}

Stream monitoring is fundamental in many data stream applications, such as financial data trackers, security, anomaly detection, and load balancing.
In that respect, quantiles are of particular interest, as they often capture the user's utility. For example, if a video connection has high tail latency, the perceived quality will suffer, even if the average and median latencies are low.

In this work, we consider the problem of approximating the \emph{per-item} quantiles. 
Elements in our stream are (ID, latency) tuples, and we wish to track the latency quantiles for each ID.
%
%
Existing quantile sketches are designed for a single number stream (e.g., containing just the latency). While one could allocate a separate sketch instance for each ID, this may require an infeasible amount of memory.
Instead, we consider tracking the quantiles for the heavy hitters (most frequent items), which are often considered particularly \mbox{important, without knowing them beforehand.}

We first present a simple sampling algorithm that serves as a benchmark. Then, we design an algorithm that augments a quantile sketch within each entry of a heavy hitter algorithm, resulting in similar space complexity but with a deterministic error guarantee.
Finally, we present \SSwRwS, a method that combines sampling and sketching while improving the asymptotic space complexity. Intuitively, \SSwRwS uses a background sampling process to capture the behaviour of the latencies of an item before it is allocated with a sketch, thereby allowing us to use fewer samples and sketches.
Our solutions are rigorously analyzed, and we demonstrate the superiority of our approach using extensive simulations.



\end{abstract}

\maketitle

\section{Introduction}
\label{sec:intro}

Maintaining statistics about network traffic is important for supporting various functionalities such as security and anomaly detection, traffic engineering, and load balancing~\cite{kabbani2010af, mukherjee1994network,garcia2009anomaly,dittmann2002network}.
Latency is an important metric in assessing a network's health and in debugging various networking middle-boxes and smart data-planes~\cite{narayana2017language}.

In particular, latency distribution is a fundamental task of data monitoring and analysis.
Yet, latency distribution is often very heavy tailed, implying that tail latency is usually more significant than average latency.
Generally, quantiles are the most commonly used for data distribution representation.
They are equivalent to the cumulative distribution function (cdf), from which the probability distribution function is derived (pdf). Thus, quantile computation is undoubtedly one of the most basic data analysis challenges.
For example, consider a large e-commerce web site that offers short average response times, but with a tail latency of several seconds, beyond what Internet users are willing to tolerate.
A customer is likely to ditch this website due to a single long latency, so having a short average response time is not enough here.

To support fast and efficient tail latency tracking, several quantile sketches have been developed~\cite{greenwald2001space, manku1998approximate, shrivastava2004medians, agarwal2013mergeable, luo2016quantiles, felber2015randomized, karnin2016optimal}.
Other research focuses on the problem's variants and extensions, such as calculating quantiles across sliding windows~\cite{arasu2004approximate}, over distributed data ~\cite{Mergeable, greenwald2004power, huang2011sampling, shrivastava2004medians}, quantile computations using GPUs~\cite{govindaraju2005fast}, continuous \mbox{monitoring of
quantiles~\cite{cormode2005holistic, yi2013optimal} and biased quantiles~\cite{cormode2006space}.}

Such sketches return an approximation of a $q$-quantile's latency up to a $\varepsilon$ error guarantee.
Existing sketches of this type track the tail latency of an entire stream.

Further, the ability to perform drill-down queries, in which we examine the behavior of the system at finer and finer granularity, may also be beneficial.
One may distinguish between different \emph{item identifiers} (or simply \emph{items}) in the stream of elements.
For example, in the case of datastores, an item identifier is typically the object's key, whereas in e-commerce sites, the item identifier might be a username or an item's SKU.
In networking applications, an item identifier may be a 5-tuple consisting of the corresponding packet's source IP, source port, destination IP, destination port, and protocol; in this case, it is common to refer to items as \emph{flows}.

Tracking tail latencies for all items can provide much richer insight about the system than looking at the aggregated tail latency, and is therefore desirable.
In the e-commerce example, a user would quit the website based on its own tail latency experience, regardless of the overall tail latency, which might be much better.
However, tracking tail latencies for all items is often impractical, given that the number of items can be extremely large and the fact that the space overhead of known quantile sketches is non-trivial.
Hence, we may instead focus on tracking tail latencies for a subset of significant items.
In particular, we are interested in the subset of \emph{heavy-hitters}, which consists of all items whose associated elements consume more than a threshold $\theta$ of the overall stream.

There are two complementary reasons why focusing on heavy-hitters makes sense in the context of tail latency monitoring.
First, since each heavy-hitter accounts for a significant fraction of the overall system load, it is important to ensure good quality of service for them.
Second, when there are only a few elements associated with a given item, e.g., the item only appears in one or two transactions, it is enough that a single transaction suffers from a longer than usual delay in order for the tail-latency of that item to be very large.
Such one-time events can be caused by, e.g., caching initialization, storage warm-up, route discovery overheads, and ``bad luck'' in terms of temporal overloads on intermediate components and devices.
On the other hand, a large tail latency for a heavy hitter points to a repetitive problem, which hopefully is easier to discover and fix, and one that is also very important to resolve.

In this work, we focus on the problem of reporting the tail latencies of heavy hitter items.
Our goal is to figure out how to find all heavy hitters' $q$-tail latencies with a maximum accuracy error of $\varepsilon$ for given parameters $q$, $\theta$ and $\varepsilon$.

\ \\
\noindent\textbf{Contributions:}

Our first contribution is the formal definition of the heavy hitters $q$-tail latencies problem, nicknamed \HHLproblem{}, where given a stream of elements and parameters $q$, $\theta$, and $\varepsilon$, we report the $q$-tail latency of every item $x$ which is larger than $\theta N$, denoted by $\widehat{\ell_x}$, such that $|\widehat{\ell_x} - \ell_x| < \varepsilon$, where $\ell_x$ is the true $q$-tail latency of $x$.

Our second contribution is a pure sampling based solution to the heavy hitters $q$-tail latencies problem called \RwS{}.
This algorithm is memory wasteful but it serves as a baseline for comparing our more sophisticated solutions.
We formally analyze this solution and show that it takes $O(\theta^{-1} \epsilon^{-2} \log\delta^{-1})$ space.

Our third contribution in a solution, nicknamed \SSwGK{}, for the heavy hitters $q$-tail latencies problem (\HHLproblem{}) which is based on combining the space saving algorithm for heavy hitters detection~\cite{SpaceSavings} with a $q$-quantile sketch.
Space saving~\cite{SpaceSavings} is considered the best approximate solution for the heavy hitters problem, as it requires $O(\varepsilon^{-1})$ space for solving the $\theta\varepsilon$ heavy hitters problem.
Formally, given user specified $\theta$, where $0 \le \theta \le 1$, a heavy hitter element is one with a frequency greater than $\theta N$ in a stream of size $N$.
In \SSwGK{}, we take the space saving data structure and add to each counter a $q$-quantile sketch, GK-sketch~\cite{greenwald2001space}.
We formally analyze this solution and show that it takes $O\parentheses{\theta^{-1}\epsilon^{-2}\cdot \log (N\epsilon^2\theta)}$ space which results in
similar space complexity as \RwS{} but with a deterministic error guarantee rather than probabilistic one.

\begin{figure*}[t]
\includegraphics[width = \linewidth]
			{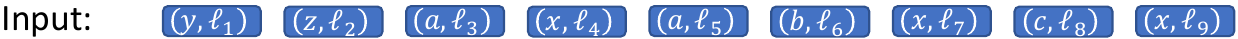}
	\begin{tabular}{ccc}
		\subfloat[\RwS{}]{\includegraphics[width = .079\linewidth]
			{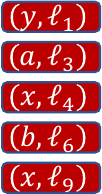}\label{fig:SQUARE}} &
		\subfloat[\SSwGK{}]{\includegraphics[width = .43\linewidth]
			{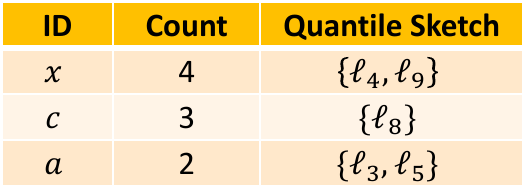}\label{fig:QUASI}} &
		\subfloat[\SSwRwS{}]{\includegraphics[width = .445\linewidth]
			{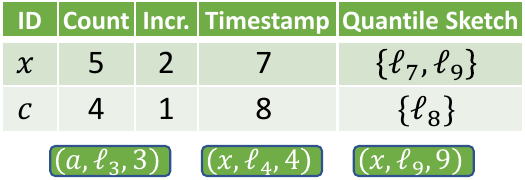}\label{fig:SQUAD}}
\\
	\end{tabular}
	\caption{An illustration of our algorithms. \RwS{} simply selects a uniform random element subset from the input stream and uses the sample to infer frequencies and quantiles. \SSwGK{} uses a heavy hitters algorithm that has a quantile sketch embedded in each counter. \SSwRwS{} combines the two approaches to asymptotically reduce the space complexity. Crucially, \SSwRwS{} adds timestamps to both samples and sketches so that it can combine an item's sketch only with the samples that were not inserted into it.
	It also adds an increments counter, which is the increase in count since the item became monitored, thus reducing the frequency estimation error. \SSwRwS{} continues to sample elements of monitored items ($(x,\ell_9,9)$ in this example) as the item may stop being monitored, i.e., the \mbox{sampling process is independent of the sketching.}}
	\label{fig:overview}
\end{figure*}

Our fourth contribution is the \SSwRwS{} algorithm, which is a combination of our first two solutions.
Figure~\ref{fig:overview} illustrates the algorithms presented in this paper.
We formally analyze this solution and show that it takes $O\parentheses{\theta^{-1}\epsilon^{-1.5}\cdot \log\epsilon^{-1}}$ space.

We present formal correctness proofs as well as space analysis.
The asymptotic space requirements of our solutions are summarized in Table~\ref{tbl:space}.
Following that, we discuss several enhancements that help our algorithms process elements more efficiently.

Our next contribution is a performance evaluation study
of the above three solutions.
To our knowledge, this is the first research to solve quantiles on a per-element level.
Thus, we compare our algorithms along with ($i$) GK-algorithm and Random~\cite{luo2016quantiles}, that serve as a best case reference point since it solves a more straightforward problem : tail latency of an entire stream ($ii$) the state-of-the-art Space Saving (SS)~\cite{SpaceSavings}, which solves the heavy hitters problem which is a building block in \SSwGK{} and \SSwRwS{}.
We evaluate our algorithms using large-scale NS3 simulations~\cite{ns3} and a FatTree topology.
The traffic is  produced using the flow size distribution in web search from Microsoft~\cite{alizadeh2010data} and Hadoop from Facebook~\cite{roy2015inside}.

The results show that given the same error guarantees $\varepsilon, \theta$, \SSwRwS{} is the most space-efficient algorithm.
While \RwS{} is the fastest algorithm in terms of update runtime, it has a large memory cost for solving the \HHLproblem{}.
When \SSwRwS{} is compared against \SSwGK{}'s update runtime, \SSwRwS{} performs better.
Yet, optimizing \SSwRwS{} enhances its update speed, making it excellent for both the performance and memory consumption metrics.
Last, we extend our results to support tail latencies for traffic volume.
All our code is open sourced~\cite{opensource}.

\begin{table}[]
\caption{A comparison of the algorithms presented in this work, in terms of their space complexity. The $\widetilde O$ notation hides polylogarithmic factors.}
\begin{tabular}{|l||l|c|c|l|}
\hline
    \textbf{Algorithm}   & \textbf{Space}              & \textbf{Deterministic}             & \textbf{Reference}        \\\hline\hline
SQUARE & $\widetilde O(\theta^{-1}\epsilon^{-2})$       &    \xmark & Section~\ref{sec:sample}     \\\hline
QUASI  & $\widetilde O(\theta^{-1}\epsilon^{-2})$       &    \cmark & Section~\ref{sec:spacesaving}     \\\hline
SQUAD  & $\widetilde O(\theta^{-1}\epsilon^{-1.5})$     &    \xmark & Section~\ref{sec:ss_w_sample}     \\\hline
\end{tabular}
\label{tbl:space}
\end{table}

\paragraph*{\textbf{Paper roadmap:}}
We briefly survey related work in Section~\ref{sec:related}.
We state the formal model and problem statement in Section~\ref{sec:prelim}.
Our first algorithm \RwS{} is described in Section~\ref{sec:sample}.
\SSwGK{} is described and analyzed in Section~\ref{sec:spacesaving}.
The improved algorithm, \SSwRwS{} is then described in Section~\ref{sec:ss_w_sample}.
We present the optimizations that enable our algorithms to process elements faster in Section~\ref{sec:opt}.
The performance evaluation of our algorithms and their comparison to GK-algorithm, Random and SS is detailed in Section~\ref{sec:eval}.
Section~\ref{sec:extensions} discusses extensions of our work.
Finally, we conclude with a discussion in Section~\ref{sec:discussion}.

\section{Related Work}
\label{sec:related}

To the best of our knowledge, this is the first work that deals with the problem of per-element quantile estimation.
Several earlier studies on streaming quantiles consider queries to be ranks, where the algorithm must associate an item $y$ in the stream with a rank close to its true rank, defined as the number of stream elements that are smaller than or equal to $y$.
In contrast, in our study, we focus on the quantile of individual elements in streams composed of identifiers and latencies, as described in Section~\ref{sec:prelim}.
Below, we discuss prior work that has been done on solving streaming quantiles that guarantee an additive error with a constant failure probability $\delta$.

Munro and Paterson included a p-pass algorithm for obtaining accurate quantiles in their classic study~\cite{munro1980selection}.
Although not explicitly studied, the method's initial run results in a streaming approach for producing approximate quantiles using $O(\epsilon^{-1}\log^{2}(N\epsilon))$ space.
Manku, Rajagopalan, and Lindsay~\cite{manku1998approximate} extended this work by proposing a deterministic solution that stores no more than $O(\epsilon^{-1} \log N\epsilon)$ objects, assuming previous knowledge of $N$.
Though \cite{manku1999random} has the same worst-case space bound, the algorithm is empirically better.
In 2001, Greenwald and Khanna~\cite{greenwald2001space} developed a complex deterministic streaming algorithm, referred to as the GK-algorithm below, that stores $O(\epsilon^{-1} \log(N\epsilon))$ objects in the worst case.
However, their experimental work used a simplified approach for which it is not clear if the $O(\epsilon^{-1} \log(N\epsilon))$ space limit still holds.
Nonetheless, they demonstrated that their method beats Manku et al~\cite{manku1998approximate}'s approach practically.
Each of these methods is deterministic and relies on comparisons.
The GK-algorithm is often considered the best in this area, both theoretically and experimentally.
Section~\ref{sec:streamingalgs} goes over it in detail.

Shrivastava et al~\cite{shrivastava2004medians} created q-digest in 2004, which is a deterministic, fixed universe method that consumes $O(\epsilon^{-1} \log \mathcal U)$ space, where $\mathcal U$ is the universe.
This approach was developed to compute quantiles in sensor networks and is a mergeable summary~\cite{agarwal2013mergeable}, a more flexible model than streaming.
However, no further efficient fixed-universe method exists in the streaming model.
Note that the $\log \mathcal U$ and $\log N$ terms are not theoretically equivalent, and \cite{shrivastava2004medians} omitted an experimental comparison with the GK-algorithm.

Randomized methods have also been considered in the past. 
The seminal results of~\cite{vapnik2015uniform} show that a random sample of size $O(\epsilon^{-2} \log \epsilon^{-1})$ contains all quantiles with at least a constant probability within the $\epsilon$ error.
This fact was shown in~\cite{manku1998approximate} and was used to compute quantiles using a random sample fed to a deterministic algorithm.
However, since this method needs knowledge of $N$ in advance, it is not a true streaming algorithm.

Manku et al.~\cite{manku1999random} developed a randomized approach that does not require knowledge of $N$ and demonstrated that the space required is $O(\epsilon^{-1} \log^{2} \epsilon^{-1})$ factor, which may be greater or less than GK's $\log N$ factor, although neither of these algorithms has been empirically tested.

Agarwal, Cormode, Huang, Phillips, Wei, and Yi~\cite{agarwal2013mergeable} proposed a mergeable sketch with the size $O(\epsilon^{-1} \log^{1.5} \epsilon^{-1})$.
For this new, simpler approach, called Random, Luo et al~\cite{luo2016quantiles} were able to provide an improved $O(\epsilon^{-1} \log^{1.5} \epsilon^{-1})$ bound.
We refer to this algorithm as "Random" and overview it in detail below.
Felber and Ostrovsky~\cite{felber2015randomized} reduced the space complexity by using a combination of sampling and the GK-algorithm to $O(\epsilon^{-1} \log\epsilon^{-1})$.

Finally, Karnin, Lang, and Liberty~\cite{karnin2016optimal} solved the problem by developing the KLL sketch, which is an optimal $O(\epsilon^{-1})$-space solution.
The KLL sketch achieves optimal accuracy in space.
The algorithm's fundamental building component is a buffer called a compactor, which accepts an input stream of $N$ items and generates a stream of no more than $\frac{N}{2}$ items that "approximates" the input stream.
The overall KLL sketch is constructed as a series of at most $\log N$ such compactors, with each compactor's output stream acting as the input stream for the next compactor.

Several studies have attempted to provide more accurate quantile estimates for low and high rankings.
Only a few provide answers to the relative error quantiles problem (also known as the biased quantiles problem).
Gupta and Zane~\cite{gupta2003counting} presented an approach for computing relative error quantiles that saves $O(\epsilon^{-3} \log^{2} (N\epsilon))$ items and uses this to estimate the number of inversions in a list; their technique needs knowledge of the stream length, $N$.
Zhang et al.~\cite{zhang2006space} previously described an approach for storing $O(\epsilon^{-2} \log  (N\epsilon^{2}))$ items.
Cormode et al.~\cite{cormode2004diamond} devised a deterministic sketch that stores $O(\epsilon^{-1} \log (N\epsilon \log(|\mathcal{U}|))$ elements and necessitates previous knowledge of the data universe $\mathcal{U}$.
Shrivastava et al~\cite{shrivastava2004medians}'s work on additive error has influenced their approach.
Zhang and Wang~\cite{zhang2007efficient} proposed a deterministic merge-and-prune method that stores $O(\epsilon^{-1} \log^{3} (N\epsilon))$ items and is capable of performing arbitrary merges with an upper constraint on n as well as streaming updates for unknown $N$.
However, it does not address the most general case of merging without prior knowledge of $N$.
Cormode and Vesely~\cite{cormode2020tight} have shown that every deterministic comparison-based technique has a space constraint of $\Omega(\epsilon^{-1} \log (N\epsilon))$ items.

Cormode et al.~\cite{cormode2021relative} presented a relative error variation of the KLL sketch.
They achieve relative error $\epsilon$ in the randomized environment using $O(\epsilon^{-1} \log^{1.5} (N\epsilon))$ with constant failure probability by varying the sampling technique throughout the distribution and employing a hierarchy modeled after~\cite{karnin2016optimal}.

\section{Preliminaries}
\label{sec:prelim}

\subsection{Model}
Given a \emph{universe} $\mathcal U$, we consider a 2-tuple \emph{stream} (sequence of elements) $\mathcal S=\angles{(x_1,\ell_1),(x_2,\ell_2)\ldots}\in (\mathcal U \times \mathbb R)^+$.
Here, each element $(x_i,\ell_i)$ has an \emph{identifier} $x_i\in\mathcal U$, and \emph{latency} $\ell_i\in\mathbb R$.

We denote by $f_x=|\set {(x_i,\ell_i)\in \mathcal S\mid x_i=x}|$ the frequency (size) of $x$.
Its (multi-) set of latencies is denoted by $L_x=\set{\ell_i\mid (x,\ell_i)\in\mathcal S}$.

Given a \emph{quantile} $q\in[0,1]$, let $\mathcal L_{x,q}$ represent the $q^{th}$ quantile (i.e., the $\ceil{q\cdot f_x}^{th}$ largest value) of $L_x$.
The inverse operation is normalized rank, denoted by \texttt{rank}$_x(\ell)$, which returns the quantile of $\ell$ in $L_x$ (that is, \texttt{rank}$_x(\mathcal L_{x,q})=q$).

Any item $x$ with frequency $f_x\ge N\theta$ is called a heavy hitter, where $N=|\mathcal S|$ is the overall number of elements, and $\theta\in[0,1]$ is a given \emph{threshold}.

Let $\epsilon,\delta\in[0,1)$ be additional error parameters: given the parameters $\theta,\epsilon,\delta$, we consider the \HHLproblem{} that tracks the latency quantiles of heavy hitters.
Specifically, we seek algorithms that support the following operations:

\begin{itemize}
    \item {\sc Insert}$(x,\ell)$ --- process a new element $(x,\ell)$.
    \item {\sc Query}$(x, q)$ --- return a tuple $(\widehat {f_x}, \widehat {\mathcal L_{x,q}})$ satisfying:
    \begin{enumerate}
        \item $\Pr[|f_x-\widehat {f_x}| > N\epsilon ]\le \delta$. As standard in heavy hitter algorithms, we return an estimate of the item's frequency.
        \item If $f_x\ge \theta N$,
        $\Pr[|\texttt{rank}(\widehat {\mathcal L_{x,q}})-q|) > \epsilon]\le \delta$. That is, if $x$ is a heavy hitter the algorithm is likely to return an estimate whose quantile is off by no more than $\epsilon$.
    \end{enumerate}
\end{itemize}

We note that part (1) of our query response is designed to help the user understand whether the quantile estimate is reliable and a similar guarantee can be obtained by running a separate heavy hitters algorithm.
Specifically, if $\widehat f_x > N\theta (1+\epsilon)$, then $x$ is likely to be a heavy \mbox{hitter and therefore $\widehat {\mathcal L_{x,q}}$ is a credible approximation of $\mathcal L_{x,q}$.}

For ease of reference, Table~\ref{tbl:notations} includes a summary of basic notations used in this work.

\begin{table}[t]
	\centering
	\small
	\begin{tabular}{|c|p{6.6cm}|}
		\hline
		Symbol & Meaning \tabularnewline
		\hline
		$\mathcal S$ & The data stream \tabularnewline
		\hline
		$\mathcal U$ & The universe of elements \tabularnewline
		\hline
        $\mathcal R$ & The universe of latencies \tabularnewline
		\hline
		$N$ & The number of elements in the stream \tabularnewline
		\hline
		$q$ & The quantile q i.e. the $q^{th}$ largest value\tabularnewline
		\hline
        $L_x$ & The set of latencies of $\mathcal S$ with identifier $x$\tabularnewline
		\hline
        $\mathcal L_{x,q}$ & The $q^{th}$ quantile of $L_x$ \tabularnewline
		\hline
        $\widehat{\mathcal L_{x,q}}$ & an estimation of $\mathcal L_{x,q}$\tabularnewline
		\hline
        rank$_x(\ell)$ & The quantile of $\ell$ in $\mathcal S_x$ \tabularnewline
		\hline
		$f_x$ & The frequency of an element $x$ in $\mathcal  S$ \tabularnewline
		\hline
		$\widehat{f_x}$ & An estimate of $f_x$ \tabularnewline
		\hline
		$\epsilon$ & An estimate accuracy parameter \tabularnewline
		\hline
		$\delta$ & A bound on the failure probability  \tabularnewline
		\hline
		$\theta$ & The heavy hitters threshold \tabularnewline
		\hline

	\end{tabular}
	\caption{List of Symbols}
	\label{tbl:notations}
\end{table}

\subsection{Useful Streaming Algorithms}~\label{sec:streamingalgs}

In this work, we utilize the Reservoir Sampling (RS) algorithm~\cite{vitter1985random} in Section~\ref{sec:sample} as well as the Space Saving (SS) algorithm~\cite{SpaceSavings} and the GK-algorithm~\cite{greenwald2001space} in Section~\ref{sec:spacesaving}.
We overview them here.

\textbf{Reservoir sampling (RS)~\cite{vitter1985random}:} is a randomized algorithm for selecting a uniform random sample of a given size from an input stream of an unknown size without replacement in a single pass through the objects. 

The algorithm keeps a $k$-sized reservoir, which initially holds the first $k$ items of the input.
On the arrival of the $n$'th item, RS selects a uniform random integer $i\in\set{0,\ldots,n-1}$; the item overrides slot $i$ of the reservoir if $i< k$  and is \mbox{otherwise~discarded}.

\textbf{Space Saving (SS)~\cite{SpaceSavings}:} is a counter-base algorithm for (approximately) finding the most frequent items in a data stream, a.k.a. heavy hitters.
SS processes a stream of identifiers with the goal of estimating the size (frequency) of each.
SS maintains a set of $1/\epsilon$ integer counters, each with an associated ID.
When an item arrives, SS increments its counter if one exists. 
Otherwise, SS allocates the item with a minimal-valued counter before incrementing it (disassociating the previous ID).
For example, assume that the smallest counter was associated with ID $x$ and had a value of $4$; if $y$ arrives and has no counter, it will take over $x$'s counter and increment its value to $5$ (leaving $x$ without a counter). 
When queried for the frequency of an item, we return the value of its counter if it has one, or the minimal counter's value otherwise. 

If we denote the overall number of insertions processed by the algorithm by $Z$, then we have that the sum of counters equals $Z$, and thus the minimal counter is at most $Z\epsilon$. 
This ensures that the error in the SS estimate is at most $Z\epsilon$.

\textbf{The GK-algorithm (GK)~\cite{greenwald2001space}:}
is a deterministic algorithm for supporting single-pass quantile summaries of a data stream.
A quantile summary is a subset of the input data sequence that uses quantile estimations to provide approximate answers to any arbitrary quantile query.

The GK technique is based on the idea that if a sorted subset of the input stream of size $N$ can be kept so that the ranks of $v_i$ and $v_{i+1}$ are within $2N\epsilon$ of each other, then any quantile query can be answered with an error no larger than $N\epsilon$. That is, given a quantile $q$, GK can produce an estimate that satisfies $\widehat q\in [q-\epsilon,q+\epsilon]$ by finding the closest ranked element in the subset.
GK allows maintaining such a subset using $O(\frac{1}{\epsilon}\log N\epsilon)$ elements, which has recently been shown to be optimal for comparison-based deterministic algorithms~\cite{cormode2020tight}.

\textbf{The Random algorithm~\cite{luo2016quantiles}:}
is an algorithm that reports all quantiles within the specified error with constant probability.

Random separates the stream into fixed-size buffers, each of which is assigned a level.
Whenever there are two buffers at the same level, Random merges them into a buffer at one level higher, such that at any time, there is at most one buffer at any level.
Random aggregates the ranks of $x$ in all buffers to report the rank of an element $x$.
Overall, it requires keeping $O(\frac{1}{\epsilon} \log^{1.5}\frac{1}{\epsilon})$ elements to guarantee that for any $q$, its estimate would satisfy $\Pr\Big[\widehat q\in [q-\epsilon,q+\epsilon]\Big]\ge 2/3$ (and this can be amplified to $1-\delta$ using the median of $O(\log\delta^{-1})$ independent~repetitions).
If we aim for a specific quantile, then the space reduces to $O(\frac{1}{\epsilon} \log\frac{1}{\epsilon})$ per repetition. 

\begin {comment}
We consider a stream of $2$-tuple elements $\mathcal S=\angles{(x_1,\ell_1),(x_2,\ell_2)\ldots}\in (U\times \mathcal R)^+$. Here, each packet $(x_i,\ell_i)$ has a flow \emph{identifier} $x_i\in U$, where $U$ is the set of all possible identifiers (e.g., all $2^{32}$ IPv4 addresses).
The second field, $\ell_i\in\mathbb R$ is the \emph{latency} of the packet, e.g., as measured by the switch by subtracting the ingress timestamp from the egress timestamp.

Given a flow $x\in U$, let $\mathcal S_x = \angles{(x_i,\ell_i)\in \mathcal S\mid x_i=x}$ be the substream of packets with flow ID $x$. The size of the flow is denoted by $f_x=|\mathcal S_x|$.
We further denote by $L_x=\set{\ell_i\mid (x,\ell_i)\in\mathcal S_x}$ its (multi-) set of latencies.
Given a \emph{quantile} $q\in[0,1]$, let $\mathcal L_{x,q}$ denote the $q^{th}$ quantile (i.e., the $\ceil{q\cdot f_x}^{th}$ largest value) of $L_x$.
The inverse operation is denoted by \texttt{rank}$_x(\ell)$, which returns the quantile of $\ell$ in $\mathcal S_x$ (that is, \texttt{rank}$_x(\mathcal L_{x,q})=q$.

Ideally, one may wish to capture the latency quantile of all flows. 
However, since the number of flows can be exceedingly large compared with the switch's memory, we must focus on a smaller set of flows.
Namely, our paper considers tracking the latency quantiles of heavy hitters (the largest flows), a set of flows that is known to be important for many networking applications~\cite{}.

Formally, given a \emph{threshold} $\theta\in[0,1]$, any flow $x$ with frequency $f_x\ge N\theta$ is considered a heavy hitter, where $N=|\mathcal S|$ is the overall number of packets.
We consider the \HHLproblem{}, where an algorithm is required to support two operations:

\begin{itemize}
    \item {\sc Insert}$((x,\ell))$ --- process a new packet $(x,\ell)$.
    \item {\sc Query}$(x, q)$ --- return a tuple $(\widehat {f_x}, \widehat {\mathcal L_{x,q}})$ satisfying the following:
    \begin{enumerate}
        \item $\Pr[|f_x-\widehat {f_x}| > \epsilon^{1/2}\theta N]\le \delta$. That is, as standard in heavy hitter algorithms, the algorithm returns an estimate of the flow size.
        \item If $f_x\ge \theta N$,
        $\Pr[|\texttt{rank}(\widehat {\mathcal L_{x,q}})-q|) > \epsilon]\le \delta$. That is, if $x$ is a heavy hitter the algorithm is likely to return an estimate whose quantile is off by no more than $\epsilon$.
    \end{enumerate}
\end{itemize}

That is, the algorithm can detect any flow whose size is at least $(\theta(1+\sqrt\epsilon))N$.

\ \newpage

A count distinct algorithm with relative error $\gamma$ requires $O(\gamma^{-2}\log\log N + \log N)$ space.

We use sampling with probability $p$ in addition to $m$ space-saving entries with a count distinct algorithm.

Therefore, the total error is:
$$
\gamma(\theta-1/m)N+1/\sqrt p\ \cdot N/m
$$
which upper bounds:
$$
N(\gamma\theta+1/\sqrt p\ \cdot 1/m)
$$
------------------------  ------------------------  ------------------------  ---
Demanding that errors are equal:
$$
\gamma\theta=1/\sqrt p\ \cdot 1/m \implies \gamma=\theta^{-1}p^{-1/2}m^{-1}
$$
Thus the error is:
$$
N(p^{-1/2}m^{-1})
$$
------------------------  ------------------------  ------------------------  ---
We want an error of $\epsilon N$
$$
p^{-1/2}m^{-1} = \epsilon
$$
and thus
$$
p = m^{-2}\epsilon^{-2}.
$$
------------------------  ------------------------  ------------------------  ---
The memory consumption is:
\begin{multline*}
Np + m(\gamma^{-2}\log\log N + \log N) \\=
Np + m(\theta^{2}\cdot p\cdot m^{2}\log\log N + \log N) \\=
Nm^{-2}\epsilon^{-2} + m(\theta^{2}\cdot m^{-2}\epsilon^{-2}\cdot m^{2}\log\log N + \log N)\\=
Nm^{-2}\epsilon^{-2} + m(\theta^{2}\epsilon^{-2}\log\log N + \log N)
\end{multline*}

Therefore, we pick 
$$
m=\sqrt[3]{\frac{N\epsilon^{-2}}{\theta^{2}\epsilon^{-2}\log\log N + \log N}}
$$

\end{comment}
\section{The \RwS{} Algorithm}
\label{sec:sample}
Here, we present the \RwSFULL{}  (\RwS{}) algorithm for solving the \HHLproblem{} that employs RS (see~\cref{sec:streamingalgs}) as a blackbox.



Intuitively, sampling is a common technique that is useful for many applications, including quantile estimation, and will serve as a baseline for our more complex algorithms.
It is long known that for approximating a quantile from a uniformly selected subset of a number stream to within an additive-$\epsilon$ error with probability $1-\delta$, one needs to sample $\theta\parentheses{\epsilon^{-2}\log \delta^{-1}}$ elements~\cite{manku1999random}.

As a result, any heavy hitter $x$ (an ID that appears at least $N \theta$ times) must be sampled $\Omega(\epsilon^{-2} \log\delta^{-1})$ times to solve the \HHLproblem{} using sampling.

Notice that there can be at most $\theta^{-1}$ heavy hitters.
As a result, we employ RS to sample $M=\Theta(\theta^{-1} \epsilon^{-2} \log\delta^{-1})$ elements from $\mathcal S$.
This way, a given heavy hitter is sampled $\Omega(\epsilon^{-2} \log\delta^{-1})$ times with probability $1-\delta/2$, and its samples allow us to produce an appropriate quantile estimate with probability $1-\delta/2$; using the union bound, we get that the overall estimate is accurate with probability $1-\delta$.

Furthermore, by selecting $\widehat f_x$ to be $N/M$ times the number of samples (e.g., see the analysis of~\cite{basat2018network}), we can estimate the frequency of an item to within an $N\cdot \epsilon\sqrt{\theta}$ factor with probability $1-\delta$.

The following theorem states the memory consumption of
the \RwS{} algorithm.

\begin{theorem}\label{sqr:space}
	\RwS{} solves the \HHLproblem{} while requiring  $O(\theta^{-1} \epsilon^{-2} \log\delta^{-1})$ space.
\end{theorem}

\section{The \SSwGK{} Algorithm}
\label{sec:spacesaving}

Next, we offer a \textbf{deterministic} algorithm called \SSwGKFULL{} (\SSwGK{}).
Intuitively, \SSwGK{} allocates a separate GK sketch~\cite{greenwald2001space} to track the latency quantiles of each potential heavy hitter.
Because we don't know the IDs of the heavy hitters ahead of time, \SSwGK{} uses a space-saving instance with $k=2\epsilon^{-1}\theta^{-1}$ entries, where each entry has a GK sketch instance configured for error $\eps_{GK}=\epsilon/2$ in addition to its counter and ID fields.
This way, \SSwGK{} can use the space saving counter value to estimate the frequency, and use the GK sketch to approximate the latency quantile.


Whenever an item $(x, \ell)$ arrives, if $x$ has an allocated counter, \SSwGK{} increments $x$'s counter and inserts $\ell$ to the associated GK instance.
Otherwise, we replace the item that has the minimal counter value with $x$ and reset its corresponding GK instance.
Then, we insert $\ell$ to this GK instance.

Using the SS variant mentioned above, we can compute any {\sc Query}$(x, q)$ as follows.
If $x$ has an allocated SS entry, we estimate its frequency using its counter value. 
Its GK instance is then queried to estimate the $q^{th}$ quantile of $L_x$.
Otherwise, if $x$ has no allocated entry, we estimate the minimal SS counter value as (an upper bound on) its frequency and do not report the latency. Since SS deterministically guarantees that every element with a frequency larger than $N/k\le N\theta$ (i.e., in particular, every heavy hitter) will have an entry, we can satisfy the accuracy guarantees.

Algorithm~\ref{alg:ss_gk} provides a high level pseudo code of \SSwGK{} using the pseudo code of Space Saving as shown in~\cite{SpaceSavings} without implementation details.
The additions to manipulate the GK sketch instances are \mbox{highlighted in blue.
Table~\ref{tbl:ss_gk} contains a list of the used variables.}

\begin{algorithm}[]

    \caption{\SSwGK{}\label{alg:ss_gk}}
	\begin{algorithmic}[1]

		\Function {Insert}{$x, \ell$}
			\If {$x$ is monitored}
				\State Increment $count_x$, the counter of $x$
				\State \textcolor{blue}{Insert $\ell$ to $GK_{x}$, the GK sketch of $x$}
			\Else
			    \If {Less than $k$ items are monitored}
    				\State $count_x \gets 1$
    				\State \textcolor{blue}{Initialize a GK sketch for $x$}    				
			    \Else			
    				\State Let $x'$ be the element with smallest $count_{x'}$
    				\State Start monitoring $x$ instead of $x'$;
    				\State $count_x \gets count_{x'} + 1$
    				\State \textcolor{blue}{Reset the GK sketch for $x$}
			    \EndIf
			\State \textcolor{blue}{Insert $\ell$ to the GK sketch of $x$}
			\EndIf
		\EndFunction
		
		\Statex
		
		\Function {query}{$x,q$}
			\If {$x$ is monitored}
				\State \Return $(count_x, \textcolor{blue}{GK_{x}.Quantile(q)})$
			\Else
				\State \Return $(count_{min}, \textcolor{blue}{undefined})$
			\EndIf
		\EndFunction
	\end{algorithmic}
\end{algorithm}

\begin{table}[]
	\centering
	\begin{tabular}{|c|p{6.6cm}|}
		\hline
		$k$ & number of entries in the SS \tabularnewline
		\hline
		$count_x$ & counter of $x$ in the SS \tabularnewline
		\hline
		$count_{min}$ & minimal counter value in the SS \tabularnewline
		\hline
		$GK_x$ & the GK sketch instance of $x$ \tabularnewline
		\hline
	\end{tabular}
	\caption{Variables used by \SSwGK{} (Algorithm~\ref{alg:ss_gk})}
	\label{tbl:ss_gk}
\end{table}


\paragraph{\textbf{Accuracy Guarantees.}\quad}
Using the standard analysis for an SS instance with k entries, we have every heavy hitter receive a space-saving instance no later than its $N/k$ arrival (because the counters sum to at most $N$, the minimal cannot be greater than $N/k$).

Therefore, if the queried element has no counter, it cannot be a heavy hitter.
Otherwise, the GK sketch of the queried heavy hitter $x$ processes all but at most $N/k=N\theta\epsilon$ latencies from $L_x$. 
Let $L_x'\subseteq L_x$ denote the subset of latencies processed by $GK_x$. 
Due to $GK_x$'s guarantees, our output $\widehat q=GK_{x}.Quantile(q)$ is within $\epsilon$ from the true quantile of $L_x'$. 
That is, $\widehat q$ deviates from the true quantile by at most $|L_x'|\cdot \eps_{GK}=|L_x'|\cdot \epsilon/2$ values.
Together with the missing latencies of $L_x\setminus L_x'$, we have that $\widehat q$ deviates by at most $|L_x'|\cdot \epsilon/2+N\theta\epsilon/2$.
In terms of quantiles, this means that our error~is 
$$
\frac{|L_x'|\cdot \epsilon+N\theta\epsilon}{|L_x|}
\le
\epsilon\cdot \parentheses{1+\frac{N\theta}{|L_x|}}
\le
\epsilon.
$$

Let us analyze the space next.
We use $k=O(\epsilon^{-1}\theta^{-1})$ entries, each with a GK sketch configured for $\eps_{GK}=\epsilon/2$.
Let $a_i$ be the number of times the $i$'th $GK$ instance was incremented; therefore, $\sum_{i=1}^k a_i\le N$.
By the space complexity of the GK algorithm we have that \SSwGK{}'s overall space requirement is
\begin{multline*}
\sum_{i=1}^k O(\epsilon^{-1}(1+\log (a_i\epsilon))) = O\parentheses{k\cdot \epsilon^{-1}\cdot\parentheses{1+ \log \frac{N\cdot \epsilon}{k}}} \\=
O\parentheses{\theta^{-1}\epsilon^{-2}\cdot \parentheses{1+\log (N\epsilon^2\theta)}}.
\end{multline*}
Here, we used Jansen's inequality and the concaveness of the logarithm function.
We summarize the analysis in the \mbox{following~theorem}.

\begin{theorem}\label{SSwGK:space}
	\SSwGK{} solves \HHLproblemParams{\theta}{\epsilon}{0} (i.e., \emph{deterministically}) while requiring  $O\parentheses{\theta^{-1}\epsilon^{-2}\cdot \parentheses{1+\log (N\epsilon^2\theta)}}$ space.
\end{theorem}

\section{The \SSwRwS{} Algorithm}
\label{sec:ss_w_sample}

The quadratic dependency on $1/\epsilon$ of the previous algorithms is sometimes prohibitively costly.
Interestingly, while both the sampling (\RwS{}) and sketching (\SSwGK{}) approaches require $\widetilde\Omega(1/\epsilon^2)$ space\footnote{The $\widetilde \Omega, \widetilde\Theta$ and $\widetilde O$ notations assume that the heavy hitters parameter $\theta$ is constant and hide polylogarithmic factors.}, applying them in tandem we can significantly improve the space complexity.
Specifically, we present \SSwRwSFULL{} (\SSwRwS{}), a hybrid algorithm that requires only $\widetilde O(\epsilon^{-1.5})$ space\addtocounter{footnote}{-1}\footnotemark{}.
Intuitively, the sampling helps us capture the behavior of the latencies experienced by an item before it was allocated with an SS entry and a quantile sketch.




\SSwRwS{} keeps $z=O(\epsilon^{-1.5}\theta^{-1}\log{\delta^{-1}})$ samples chosen by RS (see~\cref{sec:streamingalgs}). Each sample is a \emph{triplet} $(ID, latency, timestamp)$. That is, when sampling an element $(x_i,\ell_i)$ we store the triplet $(x_i,\ell_i,i)$.
Additionally, \SSwRwS{} employs an enhanced Space Saving~\cite{SpaceSavings} (SS) as described in \SSwGK{} (Section~\ref{sec:spacesaving}), but instead of using instances of the GK algorithm, it employs instances of the Random algorithm (configured for $\eps/2$ error similarly to \SSwGK{})\footnote{We chose Random as it is the fastest algorithm we are aware of, and our guarantee is random anyhow due to the sampling. One can replace it with a state-of-the-art sketch such as KLL~\cite{karnin2016optimal}, slightly improving the accuracy at a potential loss of speed.
In any case, the complexity remains $\widetilde\Theta(\epsilon^{-1.5})$.}

In contrast to~\SSwGK{}, which requires $k=\Theta(\epsilon^{-1}\theta^{-1})$ entries, \SSwRwS{} only uses $m=4\epsilon^{-0.5}\theta^{-1}$.
Intuitively, we can avoid sketching additional latencies from $L_x$ because the ones processed before $x$ is allocated to a sketch, are well approximated by the sample.

In addition to the counter and the Random instance, an entry for ID $x$ in the SS structure has a timestamp ($t_x$) that indicates when $x$ was last allocated with an entry.

If RS decides to consider the $i$'th element $(x_i, \ell_i)$, we store $(x_i, \ell_i, i)$ in the samples array.
After that, we update the augmented SS as follows:
If $x_i$ has an allocated counter, \SSwRwS{} increments it and inserts $\ell_i$ into the associated random instance.
Otherwise, we reallocate the counter with the lowest value for $x$, flush its Random instance, and set its $t_{x_i}$ to $i$.
After that, we add $\ell$ to this Random instance.
Notice that $x$ continues to participate in the RS process \emph{regardless of whether it has a counter in SS or not.}
Intuitively, $x$'s entry could become minimal and it could be evicted from the SS, so we keep tracking it by sampling.

For answering {\sc Query}$(x, q)$, we search for $x$ in the augmented SS.
If $x$ does not have an entry, our algorithms cannot promise anything about its $q^{th}$ quantile (similarly to \SSwGK{}, this means that $x$ is not a heavy hitter).
To estimate the frequency of an item $x$, we use \emph{both the sample and the SS counter}.
Specifically, let $t_x$ denote the timestamp of $x$ in the SS, and let $S_x$ denote the number of samples that belong to $x$ with a timestamp smaller than $t_x$.
We estimate the number of times that $x$ arrived \emph{before} $t_x$ as $N/z\cdot S_x$ because the probability that RS samples a specific element is $z/N$.
As a result, we estimate the frequency as $\widehat f_x = N/z\cdot S_x + I_x$.


The latencies are estimated in a similar way: we take the samples collected before $t_x$ as representing the latencies before $t_x$ and merge them with the entries stored in $Random_x$ (that represent entries between $t_x$ and $N$).

To merge the samples and the sketch, we duplicate $Random_x$ and then insert the $S_x$ samples, each with a weight of $N/z$.
Our approximation of the $q$'th quantile is \mbox{the quantile of the combined array.}


An alternative approach is to merge the samples of $x$, $Samples_x$, with the buffers of the Random algorithm in side buffers in the function {\sc Query}$(x, q)$, and then report the rank of $x$ using the merged buffers.
This approach requires an additional modification in the implementation of the Random {\sc Quantile}$(q)$ function.


The variables of the \SSwRwS{} algorithm are described in Table~\ref{tbl:ss_r_s} and its pseudocode appears in Algorithm~\ref{alg:ss_r_s}.

\begin{algorithm}[]

    \caption{\SSwRwS{} \label{alg:ss_r_s}}
	\begin{algorithmic}[1]

		\Function {Insert}{$x, \ell$} 
		    \State $n \gets n + 1$ \Comment{The current timestamp}
		    \State $\ensuremath{\mathit{RS}}.Add(x, \ell, n)$ 
			\If {$x$ is monitored}
				\State Increment $count_x$, the counter of $x$
				\State Increment $I_x$, the count since $x$ became monitored
				\State Insert $\ell$ to $Rnd_{x}$, the Random sketch of $x$
			\Else
			    \If {Less than $m$ items are monitored}
    				\State Initialize a Random sketch for $x$    	
    				\State $count_x \gets 1$
			    \Else
    				\State Let $x'$ be the element with smallest $count_{x'}$
    				\State Start monitoring $x$ instead of $x'$;
    				\State $count_x \gets count_{x'} + 1$
    				\State Reset the Random sketch for $x$
				\EndIf
				\State $I_x \gets 1$
				\State $t_{x} \gets n$
				\State Insert $\ell$ to the Random sketch of $x$
			\EndIf
		\EndFunction
		\Statex
		
		\Function {query}{$x,q$}
            \State $S_x \gets 0$
            \State $SList_x \gets \textit{empty list}$

        	\For {$j \in 0,1,\ldots, z$}
			    \If {$\ensuremath{\mathit{RS}}[j].ID = x$ and $\ensuremath{\mathit{RS}}[j].ts < t_x$}
			        \State $S_x \gets S_x + 1$
    				\State Insert $\mathit{RS}[j].\ell$ to $SList_x$
			    \EndIf
		    \EndFor
            \State $\mathit{sf_x} = \frac{n}{z} \cdot S_x $
			\If {$x$ is monitored}
				\State $RndNew_{x} = Rnd_{x}$
    			\State Insert samples from $SList_x$ with weight $\frac{n}{z}$ to $RandNew_{x}$
				\State \Return $(\mathit{sf_x} + I_x, RandNew_{x}.Quantile(q))$
			\Else
				\State \Return $(\mathit{sf_x}, undefined)$
			\EndIf
		\EndFunction
	\end{algorithmic}
\end{algorithm}

\begin{table}[]
	\centering
	\begin{tabular}{|c|p{6.6cm}|}
		\hline
		$n$& number of arrived elements
		\tabularnewline
		\hline
		$z$ & samples size used by RS
		\tabularnewline
		\hline
		$RS$ & a Reservoir Sampling instance with maximal $z$ samples.
		\tabularnewline
		\hline
		$m$ & number of entries in the SS \tabularnewline
		\hline
		$count_x$ & counter of $x$ in the SS \tabularnewline
		\hline

		$I_x$ & the count since $x$ became monitored\tabularnewline
		\hline

		$t_x$ & timestamp of $x$ in the SS \tabularnewline
		\hline

		$S_x$ & number of samples that belongs to $x$ \tabularnewline
		\hline

		$\mathit{sf_x}$ & estimation of $x$'s frequency before $t_x$\tabularnewline
		\hline
		
		$Rand_x$ & the Random instance of $x$ \tabularnewline
		\hline
	\end{tabular}
	\caption{Variables used by \SSwRwS{} (Algorithm~\ref{alg:ss_r_s})}
	\label{tbl:ss_r_s}
\end{table}

\paragraph{\textbf{Accuracy Guarantees.}\quad}
Intuitively, our analysis relies on the observation that if the sample approximates $x$'s frequency before $t_x$ to within an $\alpha$ additive error, and the space saving approximates its frequency since $t_x$ to within $\beta$ elements, then the error of the merging process cannot exceed $\alpha+\beta$; a similar logic applies to the latency quantiles (e.g., see~\cite{greenwald2004power}).

Let us start by analyzing the sample.
Let $f_{x,1}$ denote $x$'s frequency before (not including) $t_x$ and let $f_{x,2}$ denote its frequency starting with $t_x$ (i.e., $f_x=f_{x,1}+f_{x,2}$).
As before, we denote by $S_x$ the number of $x$'s samples collected before $t_x$ (observe that $S_x\sim \mbox{Hypergeometric}(N,f_{x,1},z)$). 
Thus, we use the approximation $\widehat {f_{x,1}}=S_x\cdot N/z$.
Denoting $p=f_{x,1}/N$, we can use standard concentration bounds (e.g.,~\cite{serfling1974probability}) on the hypergeometric distribution to bound the sampling error as, 
for any $\Delta\in (0,z\cdot p]$
\begin{align}\label{eq:chernoff4hypergeometric}
    \Pr\brackets{\abs{S_x-\mathbb E[S_x]}\ge \Delta} < 2e^{-\frac{\Delta^2}{3z\cdot p}}.
\end{align}


Notice that once an item that reaches a frequency of $N/m$ it cannot have the minimum SS entry. Therefore, we have that $f_{x,1} \le N/m$.
As the sampling error is monotonically increasing in $f_{x,1}$ (for $f_{x,1}<N/2$), we bound the error by analyzing the error of an item with $f_{x,1}=N/m$.
In our context, we sample $z=O(\epsilon^{-1.5}\theta^{-1}\log{\delta^{-1}})$ elements from the stream; that is, the probability for each of the $z$ \mbox{samples to belong to the first $f_{x,1}$ insertions of $x$ is $p=\frac{f_{x,1}}{N}=1/m$.}

\noindent Next, let $\Delta=\sqrt{\frac{3z\log(2/\delta)}{m}}=\Theta\parentheses{\sqrt{\frac{z\log\delta^{-1}}{m}}}$
\footnote{Observe that $z\cdot p=z/m=\Theta(\epsilon^{-1}\log\delta^{-1})$, and therefore:
$$\Delta=\Theta\parentheses{\sqrt{\frac{z\log\delta^{-1}}{m}}}=\Theta\parentheses{ \sqrt{{\epsilon^{-1}}}\cdot\log\delta^{-1}}=o\parentheses{z\cdot p}.$$}.
Our goal in what follows is to show that the error in estimating $f_{x,1}$ is likely to be lower than $N\cdot\sqrt{\frac{3\log(2/\delta)}{z\cdot m}}=\Theta(N\cdot \epsilon\cdot \theta)$.

Using~\eqref{eq:chernoff4hypergeometric}, we have that:

\begin{multline*}
\Pr\brackets{|\widehat {f_{x,1}}-f_{x,1}|\ge N\cdot\sqrt{\frac{3\log(2/\delta)}{z\cdot m}}}
\\
=\Pr\brackets{|S_x\cdot N/z-\mathbb E[S_x]\cdot N/z|\ge \frac{N}{z}\cdot \Delta}
\\
=\Pr\brackets{|S_x-\mathbb E[S_x]|\ge \Delta}
\le 2e^{-\frac{\Delta^2}{3z\cdot p}}
= \delta.
\end{multline*}

Next, recall that $f_{2,x}$ is calculated accurately using $I_x$, and therefore $\abs{\widehat{f_x}-f_x}=\abs{\widehat{f_{x,1}}-f_{x,1}}$.
Therefore, we established that the frequency estimation error is bounded by $N\cdot \epsilon\cdot \theta$, with probability $1-\delta$, using $z=c\cdot \epsilon^{-1.5}\theta^{-1}\log\delta^{-1}$ samples, for an appropriate constant $c>0$.

To analyze the quantile estimation error, we consider the error of the sampling phase (before $t_x$) separately from the error once $x$ is allocated with a sketch (starting with $t_x$).
An analysis similar to the above (with different constants) yields that the error in the sampling phase is bounded by $\epsilon/2$ except with probability $\delta/2$. 
Specifically, we can get $S_x = \Omega(z\cdot f_{x,1}/N)=\Omega(\epsilon^{-1}\log\delta^{-1})$ samples except with probability $\delta/4$, and have these approximate the quantile within an additive $\Theta(\sqrt\epsilon)$ error except with further $\delta/4$ error probability. This means that the rank of the latency is off by at most $\Theta(\sqrt\epsilon\cdot f_{x,1})=\Theta(\sqrt\epsilon\cdot N/m)=\Theta(N\epsilon\theta)$ from the true quantile.
Therefore, by configuring the quantile sketch to have an $\epsilon/2$ error with probability $1-\delta/2$, we can get that the overall estimate error, which results from the combination of the sample and the sketch, is bounded by \mbox{$f_x\cdot \epsilon + \Theta(N\epsilon\theta) = O(f_x\epsilon)$, as $f_x\ge N\theta$ per our problem definition.}

\noindent We summarize the analysis in the following theorem.

\begin{theorem}\label{SSwGK:space}
	\SSwRwS{}{} solves \HHLproblem{} \emph{deterministically} while requiring  $O\parentheses{\theta^{-1}\epsilon^{-1.5}\cdot \log\epsilon^{-1}}$ space.
\end{theorem}

\section{Optimizing the Processing Speed}
\label{sec:opt}

We now detail several optimizations that enable our algorithms to process elements faster.
First, we use the Algorithm L~\cite{li1994reservoir}, which provides a fast simulation of RS.
Intuitively, instead of drawing a random integer per item, it generates geometric random variables that represent how many items to skip before the next one is admitted into the reservoir. 
Once an item is chosen, it replaces a uniform slot in $\set{0,\ldots,k-1}$.
As a result, the total number of updates falls to $O(k(1+\log (N/k))$, implying that it takes $o(1)$ computation per element because the majority are skipped.

While we can use the above to optimize the RS process, \SSwGK{}'s and \SSwRwS{}'s processing speed is limited as arrivals of elements not tracked by the SS require initializing a new sketch.
To speed up the processing of both, we propose using an initial probabilistic filtering stage.
Intuitively, as both quantiles and frequencies can be accurately estimated from sampled streams for heavy hitters, we can process a small (e.g., 10\%) of the input and obtain rather precise results.
Namely, consider a wrapper that with probability $\mathfrak p$ calls the {\sc Insert} function of \SSwRwS{} (or \SSwGK{}\footnote{Note that this makes the algorithm randomized.}) and otherwise ignores the packet.
This means that the algorithms look at a \emph{sampled stream} $\mathcal S'\subset \mathcal S$ such \mbox{that each element in $\mathcal S$ appears with probability~$\mathfrak p$ in $\mathcal S'$ i.i.d.}

Intuitively, for a constant error probability, the sampling error would be of size $\Theta(\sqrt{N\theta/\mathfrak p})$; if this is comparable or smaller than the $\Theta(N\epsilon\theta)$ error of \SSwRwS{}, we can compensate for the error resulting from analyzing $\mathcal S'$ (rather than $\mathcal S$) without asymptotically increasing the space requirements.

There are several approaches to selecting $\mathfrak p$. 
One option is to dynamically change $\mathfrak p$ as $N$ grows, inserting elements with a weight of $1/\mathfrak p$, e.g., as suggested by~\cite{basat2019black,liu2019nitrosketch}.
For simplicity, here we consider using a fixed probability, which means that the accuracy guarantees of the algorithms only hold after a short \emph{convergence time} (as common in some sampling algorithms~\cite{basat2018volumetric,ben2017constant,mcgregor2016space}).
Namely, consider setting \SSwRwS{} to solve the \HHLproblemParams{\theta}{\alpha\cdot \epsilon}{\alpha\cdot \delta} for some $\alpha\in(0,1)$ (e.g., $\alpha=0.9$). 
Then, if the frequencies and latency quantiles are maintained in $\mathcal S'$ (the frequency, after scaling by $1/\mathfrak p$) to within error $(1-\alpha)\epsilon$, except with probability $(1-\alpha)\delta$, then the overall scheme solves \HHLproblem{}.
Here, $\alpha$ is a tradeoff parameter: the larger $\alpha$ is, the less space the algorithm requires, but also the higher the sampling probability needs to be.

As analyzed above, a sample of size $|\mathcal S'|=\Omega(\theta^{-1}\epsilon_s^{-2}\log\delta_s^{-1})$ is enough for $\mathcal S'$ to be an $\epsilon_s$ approximation of the quantiles and frequency (the $_s$ subscript represents sampling) of an element except with probability $\delta_s$.
In our case, we have $|\mathcal S'|=N\mathfrak p$, i.e., $N\mathfrak p=\Omega(\theta^{-1}\epsilon_s^{-2}\log\delta_s^{-1})$, and thus we need a convergence time of at least $N=\Omega(\theta^{-1}((1-\alpha)\epsilon)^{-2}\mathfrak p^{-1}\log((1-\alpha)\delta)^{-1})$ elements before the algorithm solves the \HHLproblem{}. 

Intuitively, since in practical applications $N\gg \theta^{-1}\epsilon^{-2}\log\delta^{-1}$, we can set a large $\alpha$ value (e.g., $\alpha=0.9)$. We can then use an intermediate value for $\mathfrak p$ (e.g., $\mathfrak p\in[0.1,0.01]$) as this gives a large speed boost and lowering the sampling probability further is not as beneficial.
This way, we do not require significantly more space (about 20\% increase for $\alpha=0.9$) nor compromise the accuracy guarantees (following the short convergence time) while significantly accelerating the solution.

\section{Evaluation}
\label{sec:eval}

\subsection{Setup}

We developed a C++ prototype for each of the algorithms mentioned in this paper: \RwS{}, \SSwGK{} and \SSwRwS{}.
The \SSwGK{} and \SSwRwS{} are implemented here using Quantile Sketch~\cite{wang2013quantiles} as a building block.
Additionally, we compared our results to the GK-algorithm~\cite{greenwald2001space} and the Random algorithm~\cite{luo2016quantiles} as a general baseline, since these are the state of the art for the more basic problem of quantiles across whole data streams, rather than per-element quantiles.
To to the best of our knowledge, this is the first study that solves quantiles on a per-element level.
Furthermore, we compared to Space Saving (SS)~\cite{SpaceSavings}, since this is a building element in \SSwGK{} and \SSwRwS{}.

\subsubsection{Dataset:}

We evaluate our algorithms using NS3 simulations~\cite{ns3} for a FatTree topology comprised of 16 Core switches, 20 Agg switches, 20 ToRs, and 320 servers (16 in each rack).
Each server has a single $100$Gbps NIC and the default load is $60\%$.
Each connection between Core and Agg switches, as well as between Agg switches and ToRs, has a capacity of $400$Gbps.
The switch buffer size is 32MB.
The traffic follows the flow size distribution in web search from Microsoft~\cite{alizadeh2010data} or Hadoop from Facebook~\cite{roy2015inside}.

The evaluation was performed on an Intel(R) 3.20GHz Xeon(R) CPU E5-2667 v4 running Linux with kernel 4.4.0-71.
Each data point in all runtime measurements is shown as a $95\%$ confidence interval of $10$ runs.
Our evaluation includes only the web search trace as the Hadoop trace exhibits very similar results.

\begin{figure*}[t]
	\begin{tabular}{ccc}
		\subfloat[\RwS{} $50\%$]{\includegraphics[width = \matrixCellWidth]
			{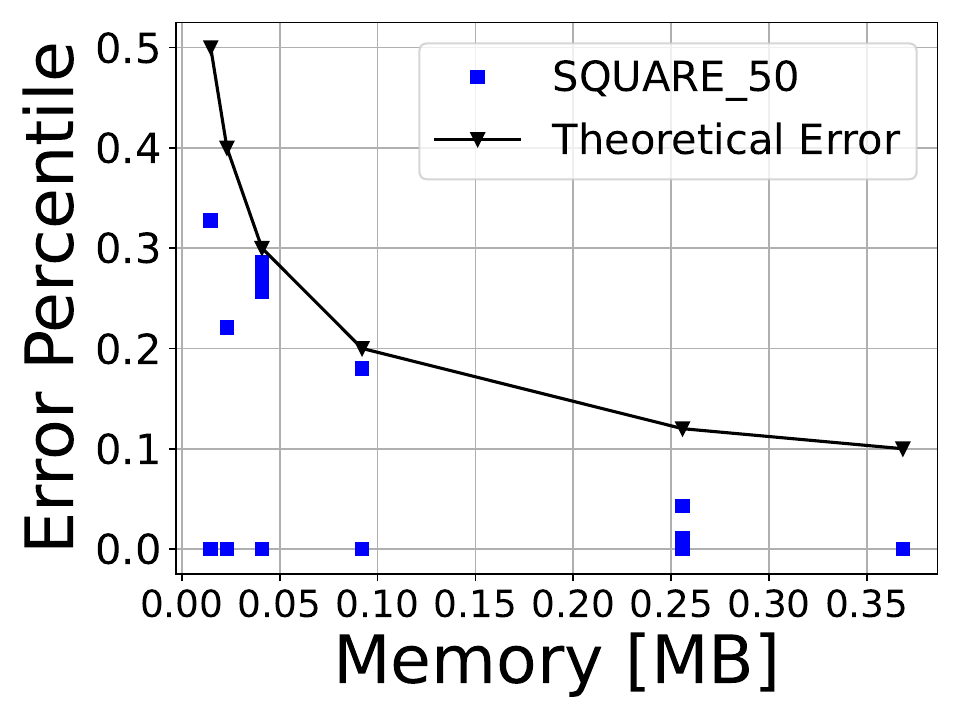}\label{fig:err_mem_square_50}} &
		\subfloat[\RwS{} $90\%$]{\includegraphics[width = \matrixCellWidth]
			{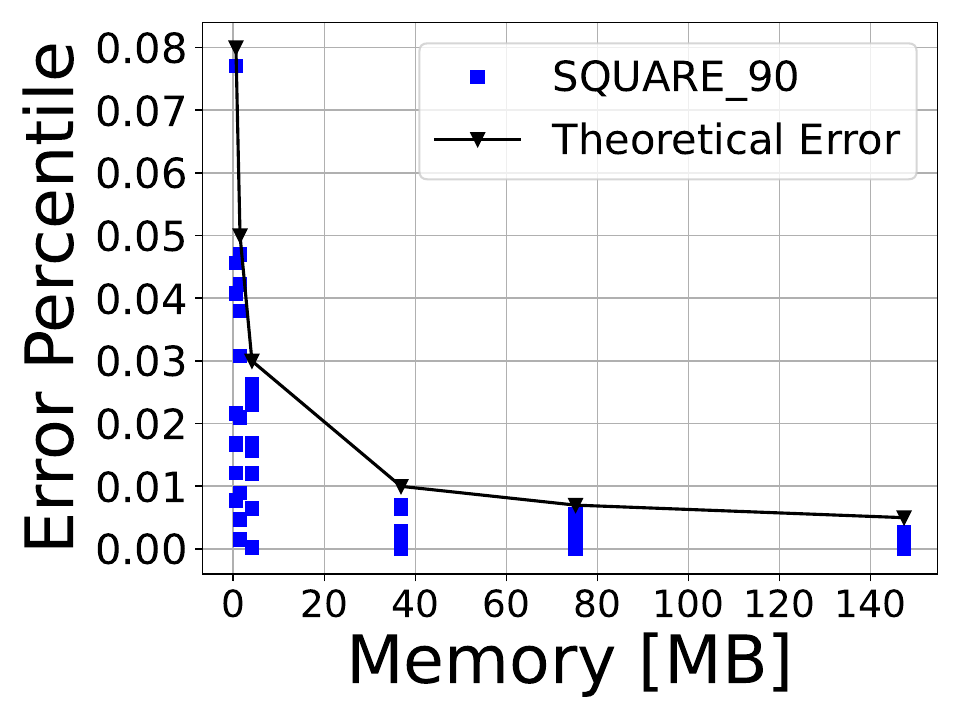}\label{fig:err_mem_square_90}} &
		\subfloat[\RwS{} $99\%$]{\includegraphics[width = \matrixCellWidth]
			{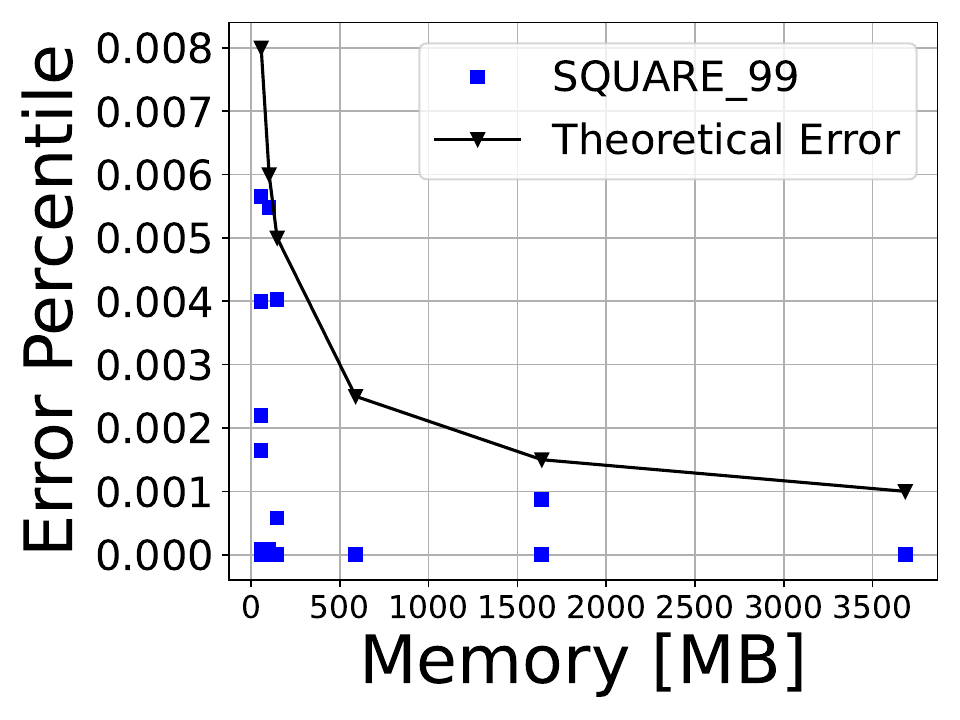}\label{fig:err_mem_square_99}}
\\
\tabularnewline
\\
		\subfloat[\SSwGK{} $50\%$]{\includegraphics[width = \matrixCellWidth]
			{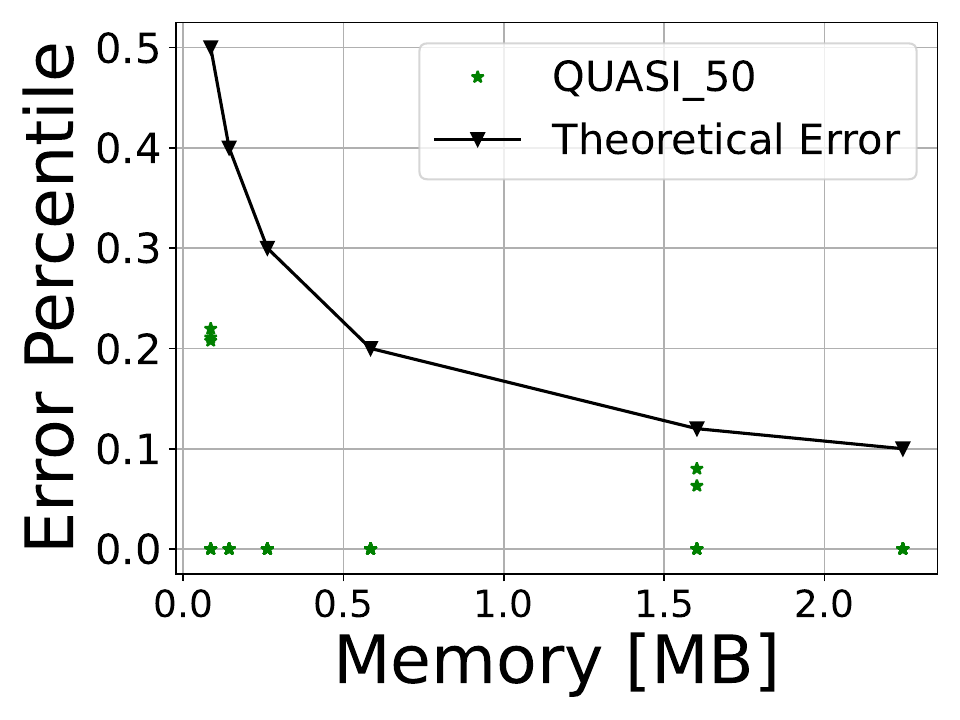}\label{fig:err_mem_quasi_50}} &
		\subfloat[\SSwGK{} $90\%$]{\includegraphics[width = \matrixCellWidth]
			{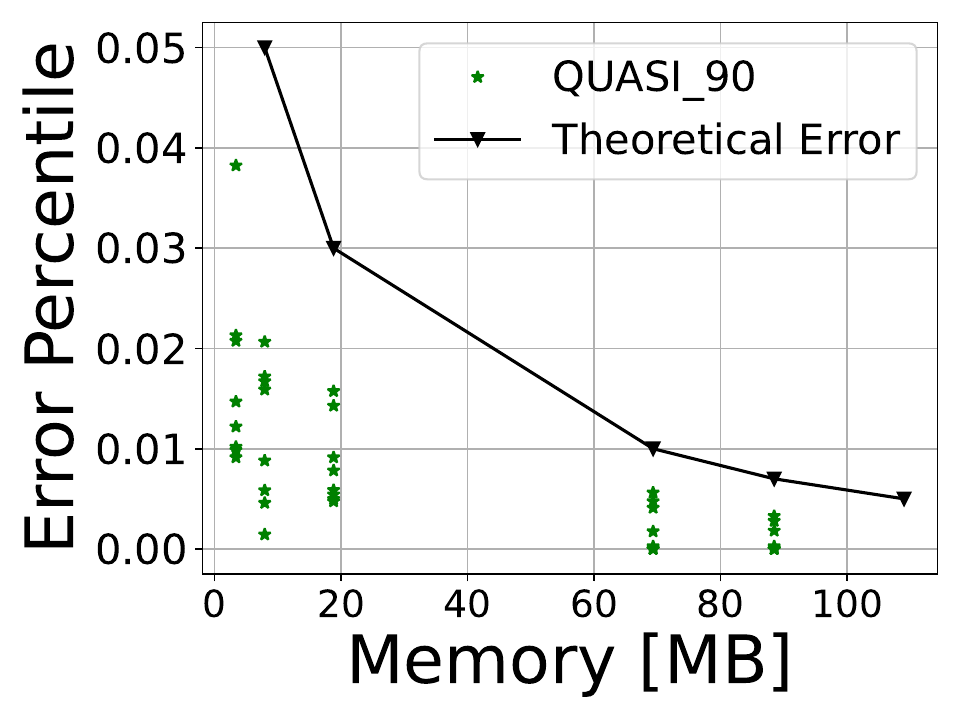}\label{fig:err_mem_quasi_90}} &
		\subfloat[\SSwGK{} $99\%$]{\includegraphics[width = \matrixCellWidth]
			{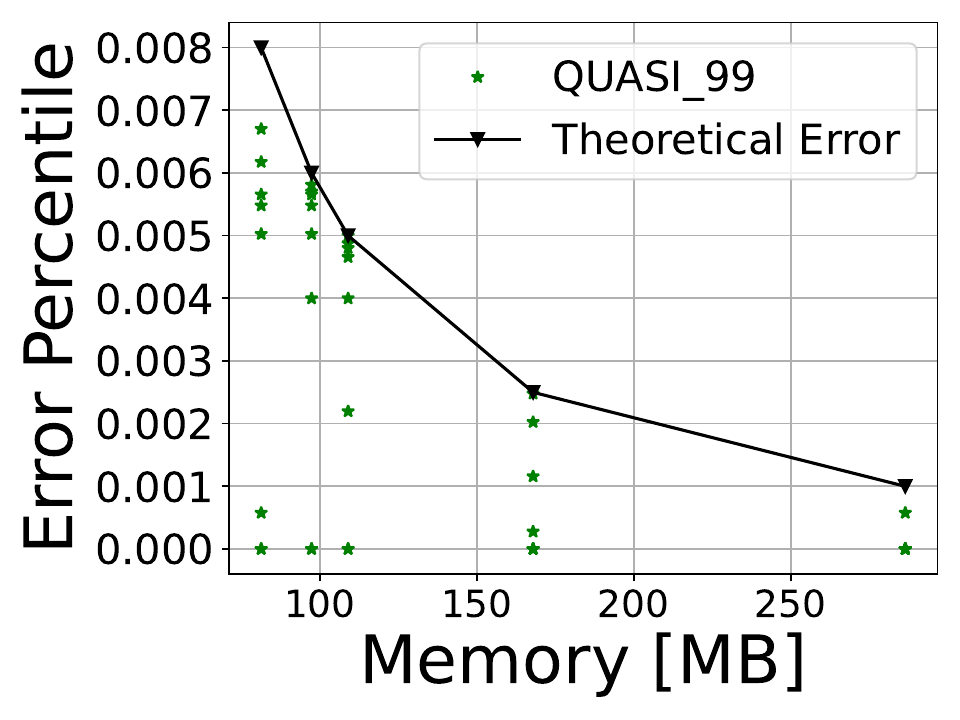}\label{fig:err_mem_quasi_99}} 
\\
\tabularnewline
\\
		\subfloat[\SSwRwS{} $50\%$]{\includegraphics[width=\matrixCellWidth]
			{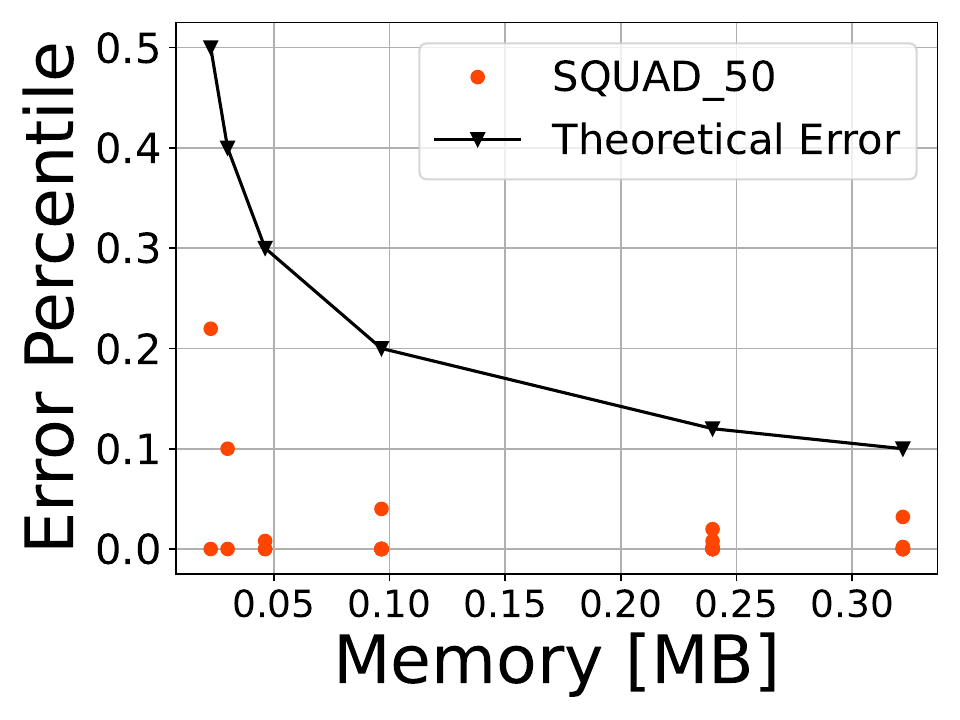}\label{fig:err_mem_squad_50}} &
		\subfloat[\SSwRwS{} $90\%$]{\includegraphics[width = \matrixCellWidth]
			{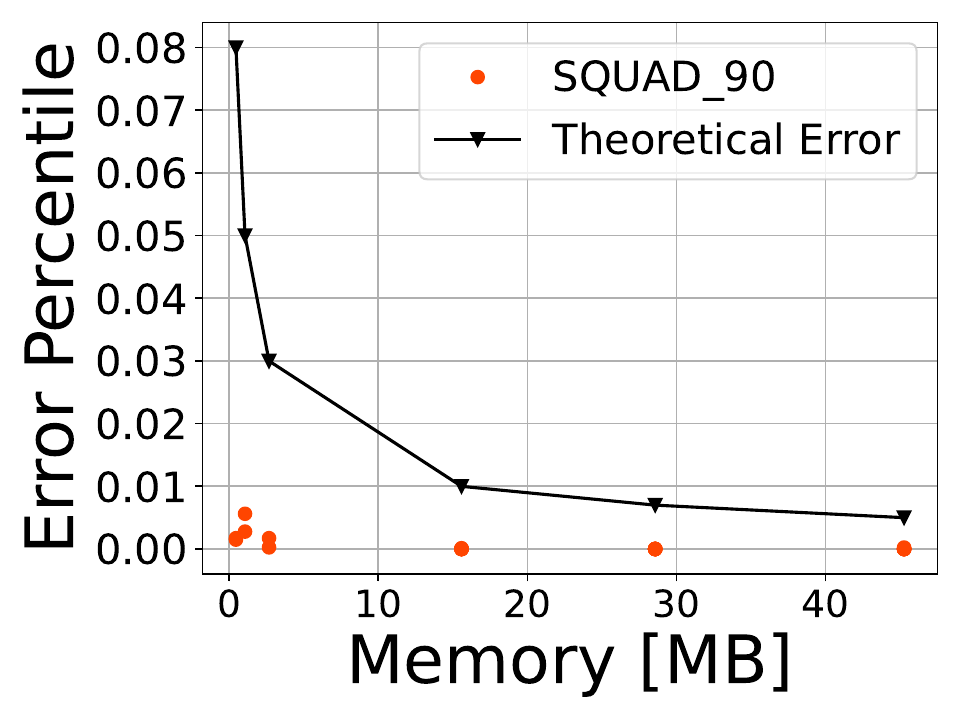}\label{fig:err_mem_squad_90}}  &
		\subfloat[\SSwRwS{} $99\%$]{\includegraphics[width = \matrixCellWidth]
			{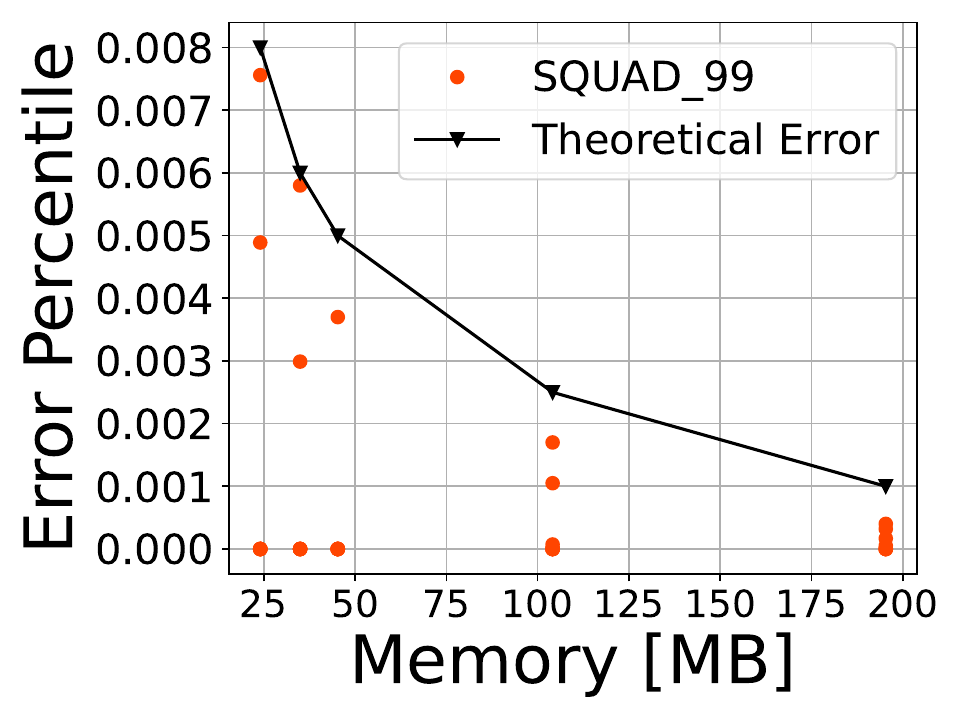}\label{fig:err_mem_squad_99}} \\
	\end{tabular}
	\caption{Accuracy as a function of used memory using NS3-simulated online search trace. Each marker corresponds with one heavy hitter; i.e., we show the error ($|\texttt{rank}(\widehat {\mathcal L_{x,q}})-q|)$) for each $x$ that satisfies $f_x\ge N\theta$ for a fixed value of $\theta = 0.01$, as a function of memory consumed. We examine the quantiles $50\%, 90\%, 99\%$ of each algorithm: \RwS{}, \SSwGK{}, and \SSwRwS{}. Notice the different x/y-axis ranges.
	}
	\label{fig:err_mem}
\end{figure*}

\begin{figure*}[t]
	\begin{tabular}{ccc}
		\subfloat[\SSwRwS{} $50\%$]{\includegraphics[width=\matrixCellWidth]
			{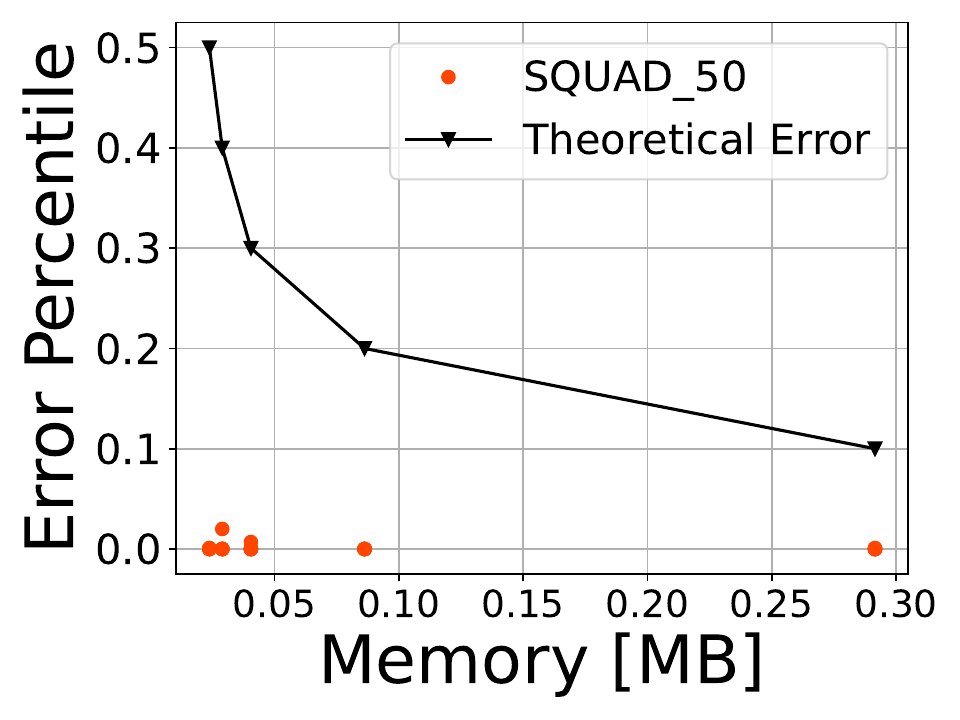}\label{fig:err_mem_squad_50}} &
		\subfloat[\SSwRwS{} $90\%$]{\includegraphics[width = \matrixCellWidth]
			{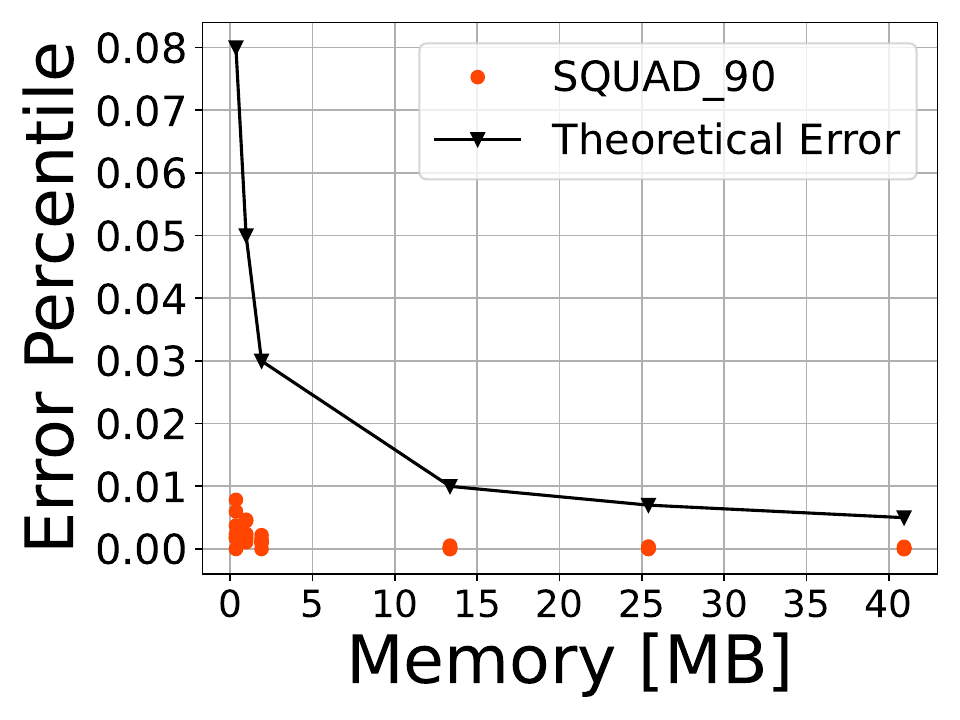}\label{fig:err_mem_squad_90}}  &
		\subfloat[\SSwRwS{} $99\%$]{\includegraphics[width = \matrixCellWidth]
			{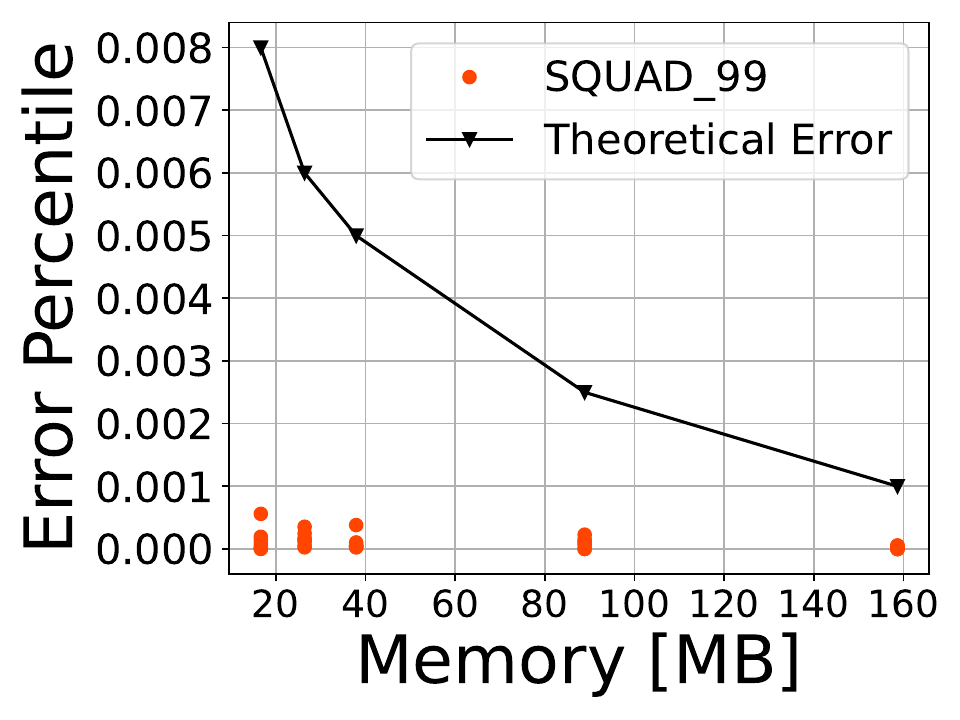}\label{fig:err_mem_squad_99}} \\
	\end{tabular}
	\caption{Accuracy as a function of used memory using NS3-simulated trace following the flow size distribution in Hadoop. Each marker corresponds with one heavy hitter; i.e., we show the percentage error ($|\texttt{rank}(\widehat {\mathcal L_{x,q}})-q|)$) for each $x$ that satisfies $f_x\ge N\theta$ with fixed value of $\theta = 0.01$, as a function of memory consumed. We examine the quantile $50\%, 90\%, 99\%$ of \SSwRwS{}.
	}
	\label{fig:FB_err_mem}
\end{figure*}

\subsection{Accuracy Comparison}
\label{sec:eval_err}
We measure accuracy in this experiment as a function of used memory.
Specifically, given quantile $q$, we measure $|\texttt{rank}(\widehat {\mathcal L_{x,q}})-q|)$, a.k.a \emph{percentage error}, as a function of consumed memory for each $x$ that satisfies $f_x\ge N\theta$.
Additionally, we present the theoretical error which demonstrates that the empirical error is constrained by the theoretical error.

Figure~\ref{fig:err_mem} illustrates the percentage error in terms of quantiles: $50\%, 90\%$ and $95\%$ for each algorithm: \RwS{}, \SSwGK{}, and \SSwRwS{} as a function of memory use with a constant value of $\theta=0.01$ using NS3-simulated online search trace.
Note that all graphs have the same amount of points, but some of them overlap in several graphs.
Additionally, Figure~\ref{fig:FB_err_mem} illustrates the percentage error for \SSwRwS{} in terms of quantiles: $50\%, 90\%$ and $95\%$ using an NS3-simulated trace following the Hadoop flow size distribution.

Throughout, as memory use increases, our algorithms get more precise, resulting in a decrease in empirical error.
As can be seen, \SSwRwS{} is the most compact algorithm among \RwS{} and \SSwGK{}, whereas \RwS{} is the most resource-intensive.
As previously stated, \RwS{} stores $\Theta(\theta^{-1} \epsilon^{-2} \log\delta^{-1})$ elements from the stream.
To ensure a small error of $epsilon$, \RwS{} should keep a high number of samples, which results in saving the whole stream size in small values of $\epsilon$ and $\theta$, as seen in Figure~\ref{fig:err_mem_squad_99}.

For the \SSwGK{} algorithms, keeping the heavy hitters in the SS instance together with their GK-algorithm sketch results in a smaller footprint than \RwS{}.
\SSwRwS{}, on the other hand, is the most efficient algorithm for solving the \HHLproblem{} due to its compact data structure, as seen in Table~\ref{tbl:space}.

In general, a lower space consumption required for a specific $\epsilon$ and $\theta$ values translates into better empirical error.
For example, \SSwGK{} consumes more memory than \SSwRwS{} for the same $\epsilon$ and $\theta$.
Thus, for a given memory budget, \SSwGK{} is more accurate than \RwS{} and \SSwRwS{} is more accurate than both.

\begin{figure*}[t]
	\center{
		\begin{tabular}{cc}
			\subfloat[\SSwGK{}]{\includegraphics[width=\matrixCellWidth]
				{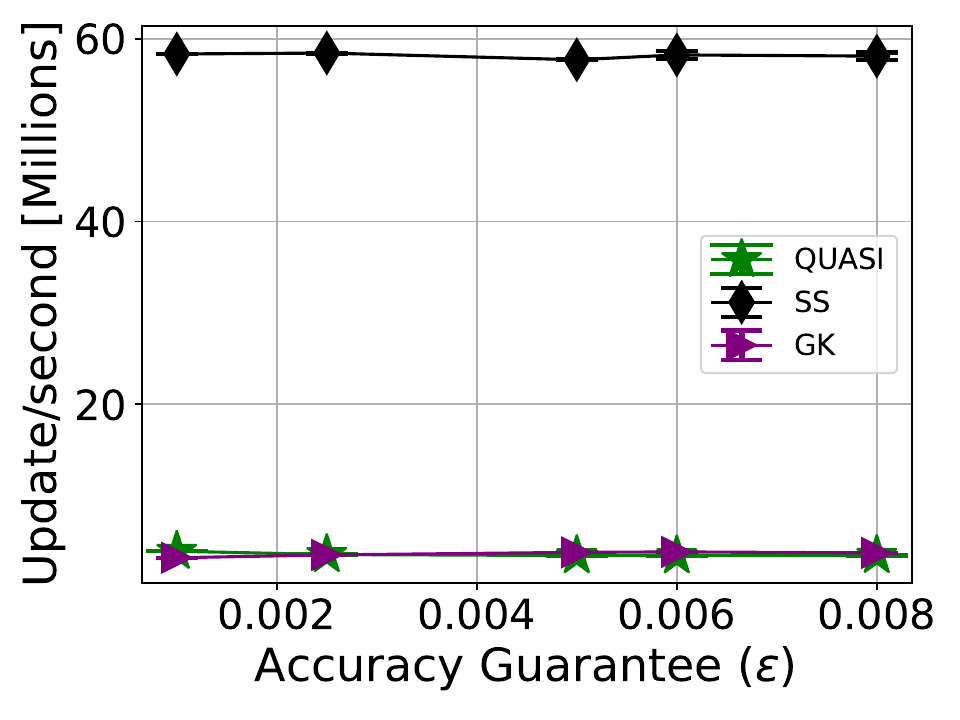}\label{fig:update_quasi}} &
			\subfloat[\SSwRwS{}]{\includegraphics[width=\matrixCellWidth]{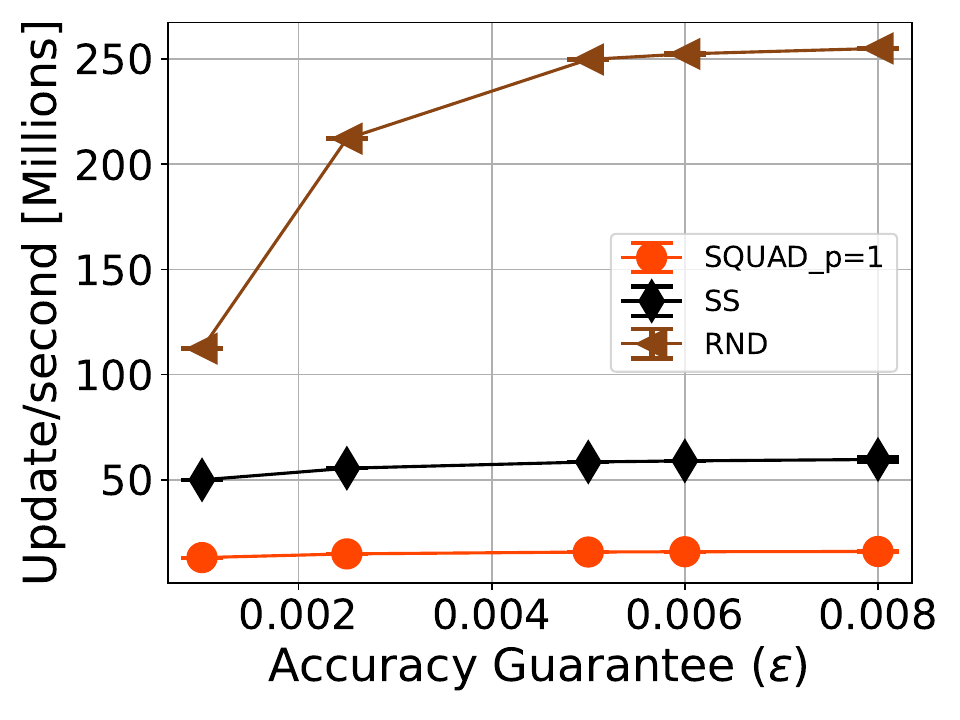}\label{fig:update_squad}}
		\end{tabular}
		}
	\caption{Update runtime as function of the accuracy guarantee ($\epsilon$) with fixed $\theta = 0.01$ (a) \SSwGK{} update performance compared to its building blocks: SS and GK (b) \SSwRwS{} update performance compared to its building blocks: SS and Random (RND in the graph). $SQUAD\_p=1$ indicates that the implementation excludes the optimizations detailed in Section~\ref{sec:opt}.}
	\label{fig:update_perf_vs_quantiles}
\end{figure*}

\begin{figure*}[t]
	\center{
		\begin{tabular}{cc}
			\subfloat[All algorithms]{\includegraphics[width=\matrixCellWidth]{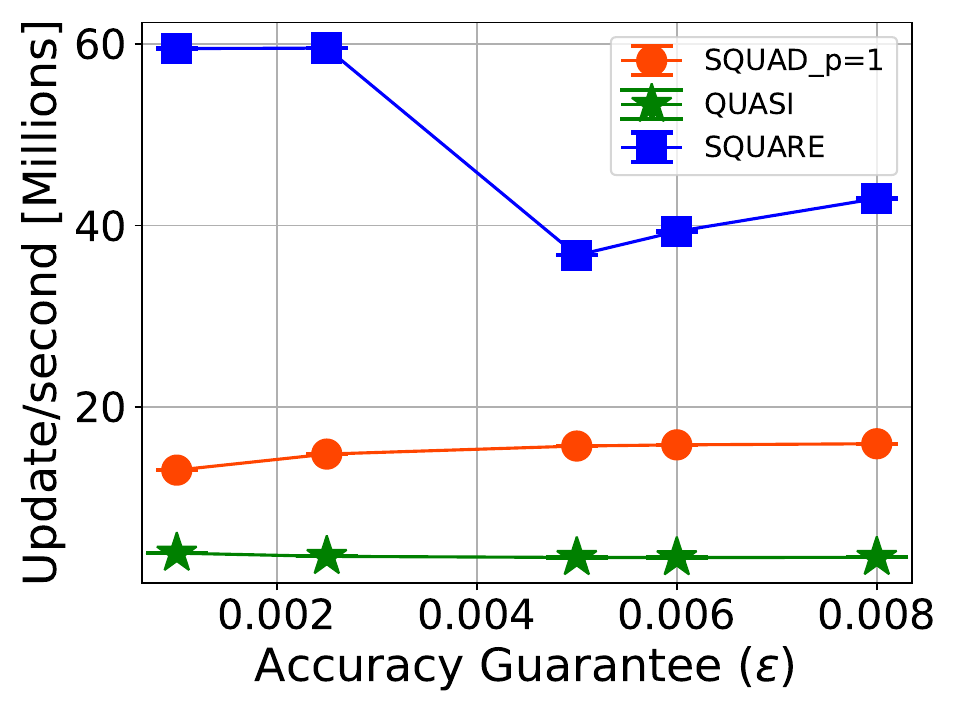}\label{fig:update_allcmp}} &
			\subfloat[Filtering]{\includegraphics[width=\matrixCellWidth]
				{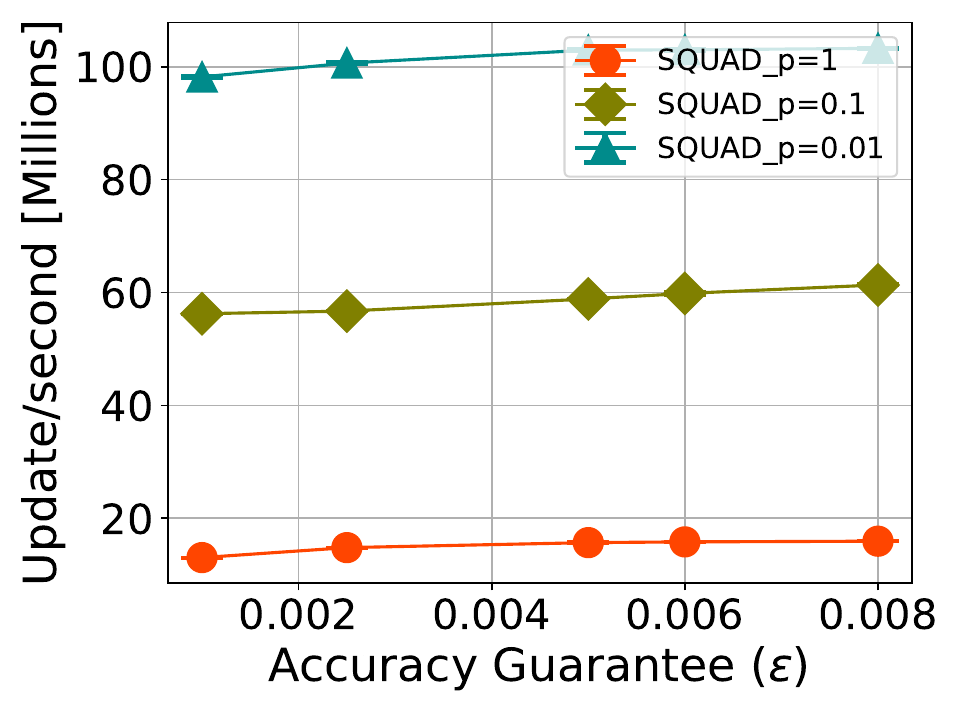}\label{fig:squad_opt}}
		\end{tabular}
		}
	\caption{Update runtime as function of the accuracy guarantee ($\epsilon$) with fixed $\theta = 0.01$ (a) Comparing all three of our algorithms: \RwS{}, \SSwGK{} and \SSwRwS{} (b) The effect of the optimization on the performance of \SSwRwS{} update.}
	\label{fig:update_perf_comparison}
\end{figure*}

\begin{figure*}[t]
	\center{
		\begin{tabular}{cc}
			\subfloat[\SSwGK{}]{\includegraphics[width=\matrixCellWidth]
				{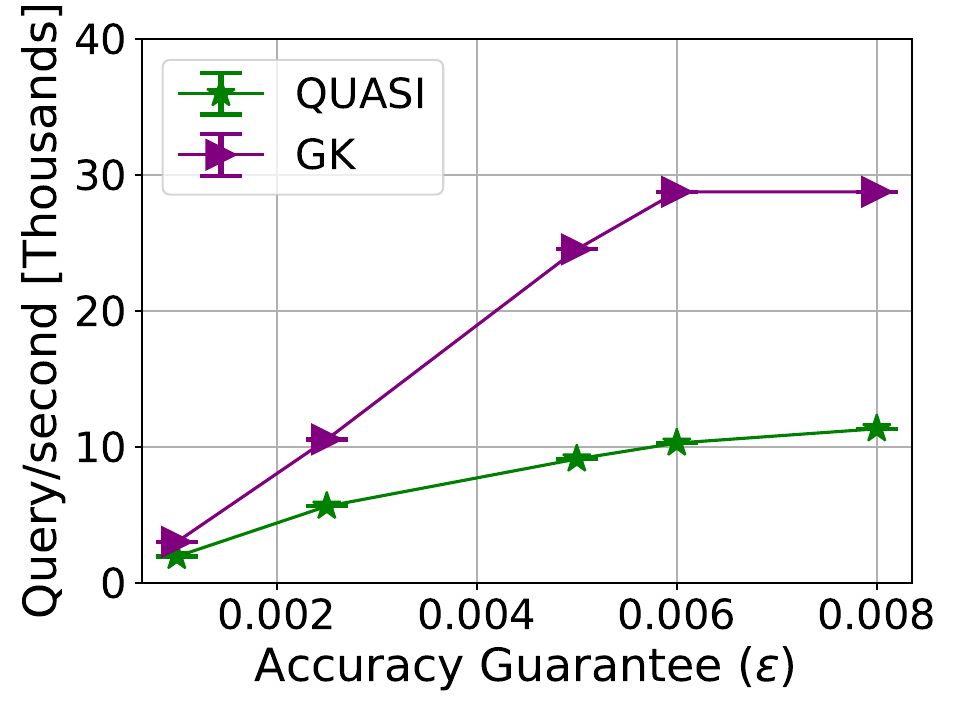}} &
			\subfloat[\SSwRwS{}]{\includegraphics[width=\matrixCellWidth]{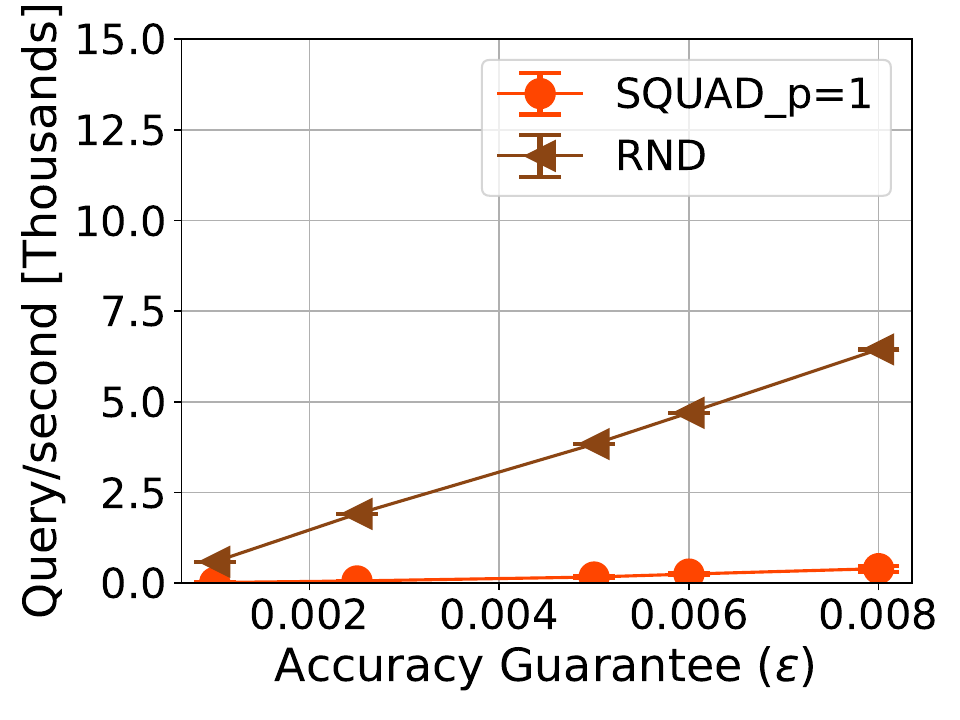}}
		\end{tabular}
		}
	\caption{Query runtime as function of the accuracy guarantee ($\epsilon$) with fixed $\theta = 0.01$ when asked for the $90 \%$ quantile (a) \SSwGK{} query performance compared to GK (b) \SSwRwS{} query performance compared to Random (RND).}
	\label{fig:query_perf}
\end{figure*}

\subsection{Performance Comparison}
\label{sec:eval_perf}

Figures~\ref{fig:update_perf_vs_quantiles} and~\ref{fig:update_perf_comparison} compare the update speed.
We explore the trade-off of $\epsilon$ with a fixed $\theta = 0.01$.

\subsubsection{\SSwGK{} Update Time:}
Figure~\ref{fig:update_quasi} illustrates the performance of \SSwGK{} in terms of update time when compared to its building blocks: Space Saving (SS)~\cite{SpaceSavings} and the GK-algorithm~\cite{greenwald2001space}.
Bear in mind that, although SS is the quickest, neither it nor the GK-algorithm solve the \HHLproblem{} and rather serve as a best-case reference point.

As can be observed, \SSwGK{}'s update performance is comparable to that of the GK algorithm.
Recall that the \SSwGK{} update operation is equivalent to an update operation in SS and an update operation on the corresponding GK instance.
Due to the high effectiveness of SS updates, the run time of \SSwGK{} updates is limited by the run time of GK.
That is, while GK-algorithm solves the quantile problem for the full stream, the \SSwGK{} algorithm solves per-element quantiles without adding any extra update time cost.

Specifically, we may replace the GK instances in \SSwGK{} with any sketch that solves quantiles, such as Random~\cite{luo2016quantiles}, which has a higher update speed, as seen in Figure~\ref{fig:update_squad}.
However, \SSwGK{} will no longer be a deterministic solution in this case.
As a result, there is a trade-off between update speed and determinism.
Additionally, it was shown that the GK-algorithm is an optimal \textbf{deterministic} comparison based algorithm.

\subsubsection{\SSwRwS{} Update Time:}
Figure~\ref{fig:update_squad} compares \SSwRwS{}'s update speed to that of its building blocks: Space Saving (SS)~\cite{SpaceSavings} and the Random-algorithms~\cite{luo2016quantiles}.

Recall that the Random algorithm reports quantiles over the entire stream, thus it does not solve the \HHLproblem{} and only serves as a best case reference point.
\SSwRwS{} update operation is translated to sampling operation, SS update and Random update.
As a result, the run time of its update is impacted by the run time of all of them.
$SQUAD\_p=1$ in the graph indicates that the implementation excludes the optimizations detailed in Section~\ref{sec:opt}.

\subsubsection{Comparing \RwS{}, \SSwGK{} and \SSwRwS{} Update Time:}
As seen in Figure~\ref{fig:update_allcmp}, \RwS{} is the fastest algorithm since each update is converted to a sampling update.
\SSwRwS{}, on the other hand, performs better than \SSwGK{} in terms of update performance since it is based on the Random algorithm, which has a faster update time than the GK- algorithm, the building block of \SSwGK{}.
While \RwS{} is the quickest, it solves the \HHLproblem{} with a high memory cost, as seen in Figure~\ref{fig:err_mem}.
When $\epsilon$ is small, \RwS{} keeps a significant number of samples more than the stream size.
In this scenario, \RwS{} just saves all streams, which results in superior performance than larger values of $\epsilon$, which allows items to override earlier samples.

Figure~\ref{fig:squad_opt} shows how the filtering optimization discussed in Section~\ref{sec:opt} improves the update performance of \SSwRwS{}.
In this scenario, when the filter sampling probability $\mathfrak p$ is $0.1$, its performance comparable or better than \RwS{}, and with $\mathfrak p=0.01$, \SSwRwS{} becomes the clear winner.
We explore these optimizations further below.

\subsubsection{Query Speed Comparison:}
\label{sec:query_op}

For comparing the query speed, we used the quantiles $50\%, 90\%, 99\%$.
We investigated the effect of the $\epsilon$ parameter using a fixed $\theta$ of $0.01$, and the experiment includes quantile queries for items $x$ that satisfy the condition $f_x\ge N\theta$.
For decreasing $\epsilon$ values, more latencies access the quantile sketches.
Consequently, we got slower query operations in all algorithm.
As seen in Figure~\ref{fig:query_perf}, \SSwGK{} performs better than \SSwRwS{} because \SSwRwS{} relies on Random queries, which perform worst than the GK.
Additionally, \SSwRwS{} checks its samples part to figure out the samples that were taken before the time the given identifier enters the SS.
This becomes extremely expensive when the sample size \mbox{is large, i.e. when the $\epsilon$ value is small.}

Particularly, as seen in Figure~\ref{fig:update_squad}, we may replace the Random instances in \SSwRwS{} with a GK sketch that has a faster query performance.
However, as seen in Figure~\ref{fig:update_perf_vs_quantiles}, GK is slower in update performance.
Indeed, there is a trade-off between update and query performance.
However, since is most streaming applications updates occur more often than query operations, Random would usually be the preferred choice.

\begin{figure*}[t]
	\center{
		\begin{tabular}{cc}
			\subfloat[Memory]{\includegraphics[width=\matrixCellWidth]
				{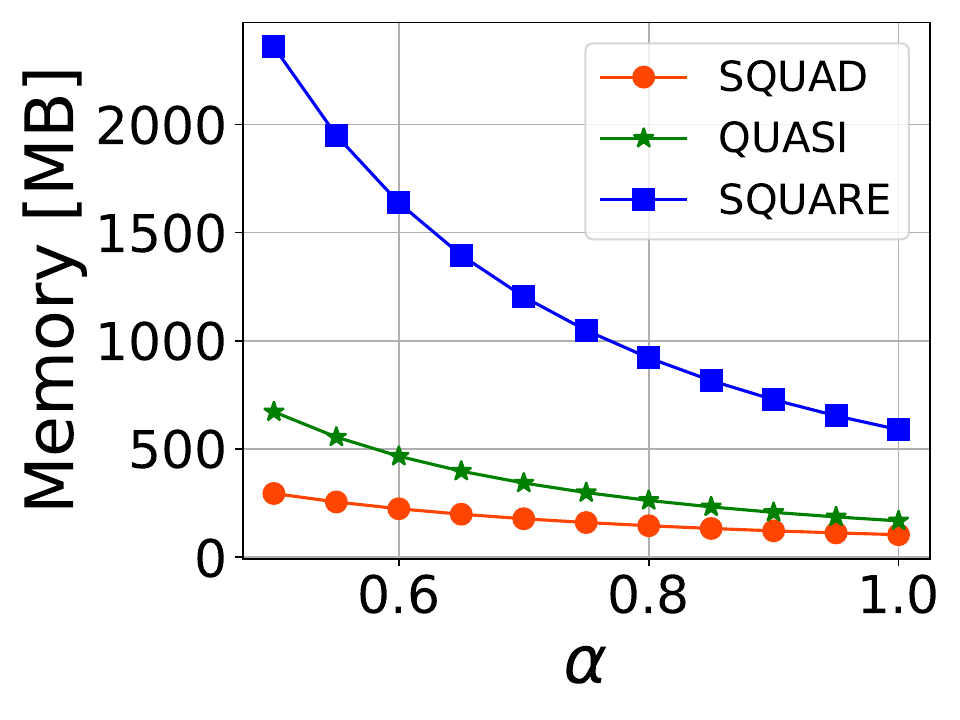}\label{fig:mem_alpha}} &
			\subfloat[Error]{\includegraphics[width=\matrixCellWidth]{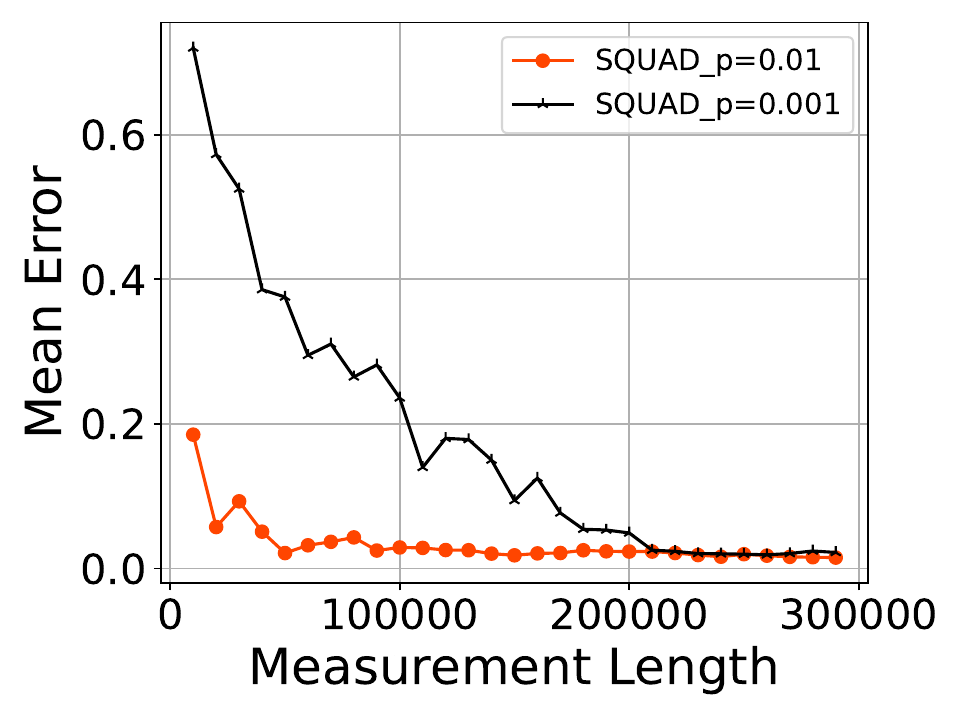}\label{fig:err_tracesize}}
			\subfloat[Error]{\includegraphics[width=\matrixCellWidth]{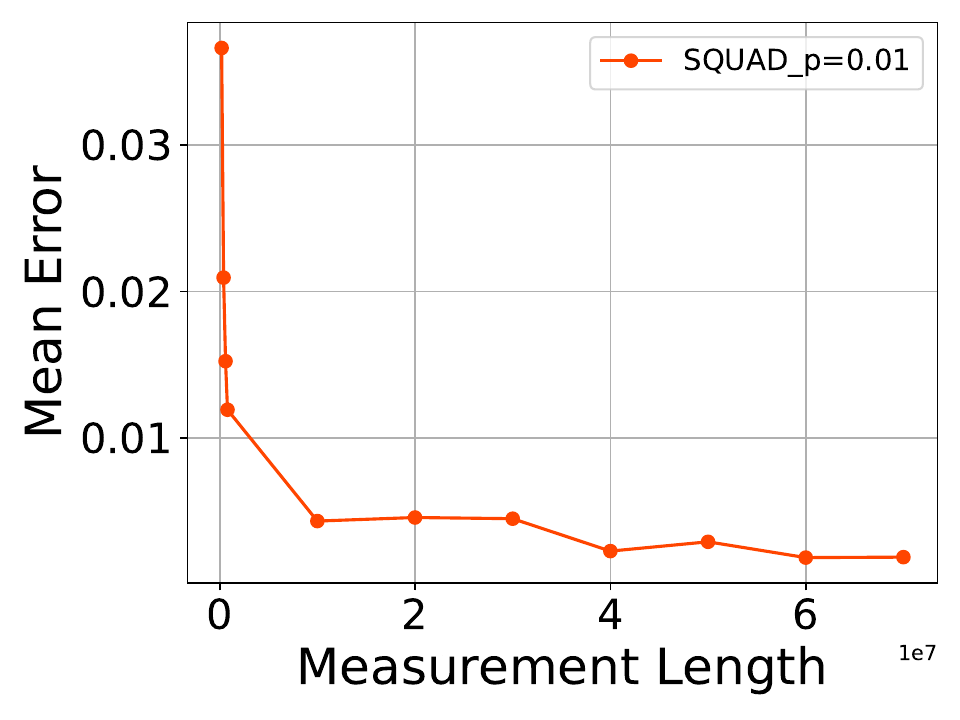}\label{fig:err_alltracesize}}

		\end{tabular}
		}
	\caption{(a) Memory consumption as function of $\alpha$ with $\epsilon=0.0025$ and $\theta=0.01$ (b) Mean Accuracy as a function of the measurement length size using an NS3-simulated trace that follows the Hadoop flow size distribution. Each marker represents the mean of the heavy hitters' percentage error, with fixed values of $\theta=0.01$ and $\epsilon=0.025$. We examine the sampling probability $\mathfrak p = 0.01, 0.001$ in \SSwRwS{} (c) Accuracy as a function of the measurement length till the stream's end when sampling with $\mathfrak p = 0.01$ in \SSwRwS{} with $\theta=0.01$ and $\epsilon=0.025$.\ran{TODO: add $p=0.1$, switch to log scale on the x-axis, merge (b) and (c), add a horizontal line for $\epsilon$.}}
	\label{fig:opt_perf}
\end{figure*}

\subsection{Optimizations Comparison}
In this section we evaluate the optimizations described in Section~\ref{sec:opt}.
We examine the influence of the optimizations on the update runtime, memory usage, and empirical error.

\subsubsection{Effect of Optimization on \SSwRwS{} Update Time:}
We implement the optimizations in \SSwRwS{} since it is the most space-efficient algorithm.
The update runtime is shown in Figure~\ref{fig:squad_opt} as a function of $\epsilon$ with a fixed $\theta=0.01$.
The three \SSwRwS{} implementations differ in the probability of the wrapper calling the {\sc Insert} function of \SSwRwS{}.
We consider three probabilities: $\mathfrak p=1$ (indicating that the implementation does not include optimizations), $\mathfrak p=0.1$, and $\mathfrak p=0.01$.
That is, each element in $\mathcal S$ occurs with probability $\mathfrak p$ in the sampled stream i.i.d.
As expected, decreasing the value of $\mathfrak p$ results in improved update speed, as the algorithm invokes the \SSwRwS{} {\sc Insert} function infrequently.
As can be observed, the optimizations considerably improve the speed of the update.

\subsubsection{Effect of $\alpha$ on Memory Consumption:}
Figure~\ref{fig:mem_alpha} shows the space consumed by our algorithms as function of $\alpha$ with fixed values of $\epsilon=0.025$ and $\theta=0.01$.
As seen in Figure~\ref{fig:mem_alpha}, the larger $\alpha$ is the less space the algorithm requires, but also the higher the sampling probability needs to be.
For $\alpha = 0.9$, with $\epsilon=0.025$ and $\theta=0.01$ we get an increase of $18\%$ in the space requirement of \SSwRwS{}.
Parameter $\alpha$ has less impact on the memory of \SSwRwS{} than its impact on \RwS{} and \SSwGK{} since \SSwRwS{} space complexity is better than the others for the same values of $\epsilon$ and $\theta$.

As seen in Figure~\ref{fig:mem_alpha}, the greater $\alpha$ is, the less space is required for the algorithm, but the sampling probability must be increased.
With $\alpha=0.9$, $\epsilon=0.025$, and $\theta=0.01$, the space needed for \SSwRwS{} increases by $18\%$.
The parameter $\alpha$ has a smaller effect on the memory of \SSwRwS{} than it does on \RwS{} and \SSwGK{}, since \SSwRwS{} has a better space complexity than the others for the same values of $\epsilon$ and $\theta$.


\subsubsection{The Effect of the Length of the Measurement on the Error:}
We study the trade-off between geo-sampling rate $\mathfrak p$ and the convergence time (in terms of the number of packets) and report the results in Figure~\ref{fig:err_tracesize}.
We use an NS3-simulated trace that follows the Hadoop flow size distribution with fixed values of $\theta=0.01$ and $\epsilon=0.025$.
Since \SSwRwS{} with optimizations uses sampling to select packets, it requires a convergence time to produce a guaranteed accurate result (analyzed in Section~\ref{sec:opt}).
As expected, larger $\mathfrak p$ value leads to faster convergence time as we sample elements in higher probability.
In addition, we examine the mean error of \SSwRwS{} during the whole trace and show it in Figure~\ref{fig:err_alltracesize}.





\section{Extensions of Supporting Tail Latencies for Traffic Volume Heavy-Hitters}
\label{sec:extensions}


It is often desirable to find the tail latencies for heavy hitters in terms of traffic volume.
That is, consider a stream in which each element has a \emph{size} and our goal is to find the tail latency for items that use the majority of the bandwidth.
Formally, we look at a \emph{weighted stream} $\mathcal S=\angles{(w_1, x_1, \ell_1),(w_2, x_2,\ell_2)\ldots}\in (\{1,2,\ldots,M\} \times \mathcal U \times \mathbb R)^+$ and define item's volume as the sum of sizes for elements that belong to it.
It is worth noting that the weight $w_i$ refers to the element $i$, which is composed of both $x_i$ and $\ell_i$.

Sampling may be performed on the basis of the number of elements or, more broadly, on the basis of some weight (e.g., the size of the associated data).
Weighted sampling provides a more precise view of the underlying byte-traffic.
This is desirable for applications such as traffic engineering and load balancing that aim to stay within the bandwidth constraints of a network, as well as writes to SSDs and corresponding write-amplification etc.
Priority Sampling~\cite{duffield2007priority} is a weighted sampling technique that is optimal. That is, when compared to other sampling techniques, Priority Sampling has a lower or equivalent variance.
The goal is to produce a sample of $k$ keys with a probability proportional to their weight.
Priority Sampling accomplishes this by assigning the value $\frac{w_i}{r}$ to each key, with $r$ randomly selected from the range $[0,1]$.
Priority sampling includes the $k$ keys with the highest values.

The Space Saving algorithm~\cite{SpaceSavings} can find weighted heavy hitters in a stream with an update time of $O(\log\epsilon^{-1})$~\cite{berinde2010space}.
Recent advancements~\cite{DIM-SUM, FAST, anderson2017high} reduce this runtime to a constant.
Thus, the tail latency problem for weighted heavy hitters may be solved with the same asymptotic complexity as the unweighted versions and with an error of up to $M\epsilon$.

Intuitively, to address this problem, we can use Priority Sampling instead of RS sampling and modify the Space Saving method to get the weighted heavy hitters.
Furthermore, because the weight $w_i$ corresponds to the latencies ($\ell_i$), we must ensure that the reporting quantile of weighted latencies is met.
KLL sketch~\cite{karnin2016optimal} can also handle weighted items.
Its size remains $O(\epsilon^{-1}\sqrt{\log\epsilon^{-1}})$ but the update time becomes $O(\log\epsilon^{-1})$.
That is, we need to replace the quantile-sketch in \SSwGK{} and \SSwRwS{} with a KLL sketch to support weighted items.

Putting it all together, in \RwS{} we employ Priority Sampling instead of RS sampling, and in \SSwGK{} we use the Space Saving version for weighted streams with KLL sketch as quantile-sketch.
In \SSwRwS{}, we employ Priority Sampling for the sampling part, as well as a Space Saving weighted variant with KLL sketch.

Thus, our algorithms are capable of solving the \HHLproblem{} for weighted streams with the same space complexity \mbox{as the unweighted version and with an error of at most $M\epsilon$.}

\section{Discussion}
\label{sec:discussion}

In this paper, we studied the problem of reporting the tail latencies of heavy hitter items.
To our knowledge, this is the first research to solve quantiles on a per-element level rather than reporting quantiles of an entire stream.
Such capabilities can be useful when one wishes to assess a network’s health and to debug various networking middle-boxes and smart data-planes.

We presented a formal definition of the generalized problem and explored three solutions: a sample approach (\RwS{}) and more sophisticated solutions called \SSwGK{} and \SSwRwS{}.
\SSwGK{} is a deterministic solution that assigns a unique quantile-sketch (GK) to each potential heavy hitter that is obtained from a Space Saving instance.
\SSwRwS{} combines the \RwS{} and \SSwGK{} algorithms, resulting in superior memory reduction.

\SSwRwS{} is the most memory-efficient algorithm.
Both \SSwRwS{} and \RwS{} use about the same amount of memory.
\SSwGK{}, on the other hand, is deterministic, but \RwS{} has an error probability.
This is true both asymptotically and in measurements throughout a large-scale NS3 simulation, where we observed orders of magnitude memory reductions for similar estimation errors in the \SSwRwS{}~algorithm.

While \RwS{} has a faster update rate than \SSwGK{} and \SSwRwS{}, it consumes a lot of memory.
To that end, we suggested several efficiency enhancements for the update operation of our algorithms in Section~\ref{sec:opt}.
In fact, the update performance of \SSwGK{} is comparable to that of the state-of-the-art method, which can only handle quantiles throughout the whole stream, not per-element quantiles.
Our approach can be applied to the case of volume traffic, where each element in the stream has a size and our algorithms determine the tail latency for the elements that use the majority of the available~bandwidth.

\textbf{Code Availability:} All code is available online~\cite{opensource}.\\

\textbf{Acknowledgements:} This work was partially funded by the Technion-HPI research school and the Israel Science Foundation grant \#3119/21.

\newpage
\bibliographystyle{abbrv}
\bibliography{refs}  

\end{document}